

\catcode`\@=11


\message{Loading jyTeX fonts...}



\font\vptrm=cmr5
\font\vptmit=cmmi5
\font\vptsy=cmsy5
\font\vptbf=cmbx5

\skewchar\vptmit='177 \skewchar\vptsy='60
\fontdimen16 \vptsy=\the\fontdimen17 \vptsy

\def\vpt{\ifmmode\err@badsizechange\else
     \@mathfontinit
     \textfont0=\vptrm  \scriptfont0=\vptrm  \scriptscriptfont0=\vptrm
     \textfont1=\vptmit \scriptfont1=\vptmit \scriptscriptfont1=\vptmit
     \textfont2=\vptsy  \scriptfont2=\vptsy  \scriptscriptfont2=\vptsy
     \textfont3=\xptex  \scriptfont3=\xptex  \scriptscriptfont3=\xptex
     \textfont\bffam=\vptbf
     \scriptfont\bffam=\vptbf
     \scriptscriptfont\bffam=\vptbf
     \@fontstyleinit
     \def\rm{\vptrm\fam=\z@}%
     \def\bf{\vptbf\fam=\bffam}%
     \def\oldstyle{\vptmit\fam=\@ne}%
     \rm\fi}


\font\viptrm=cmr6
\font\viptmit=cmmi6
\font\viptsy=cmsy6
\font\viptbf=cmbx6

\skewchar\viptmit='177 \skewchar\viptsy='60
\fontdimen16 \viptsy=\the\fontdimen17 \viptsy

\def\vipt{\ifmmode\err@badsizechange\else
     \@mathfontinit
     \textfont0=\viptrm  \scriptfont0=\vptrm  \scriptscriptfont0=\vptrm
     \textfont1=\viptmit \scriptfont1=\vptmit \scriptscriptfont1=\vptmit
     \textfont2=\viptsy  \scriptfont2=\vptsy  \scriptscriptfont2=\vptsy
     \textfont3=\xptex   \scriptfont3=\xptex  \scriptscriptfont3=\xptex
     \textfont\bffam=\viptbf
     \scriptfont\bffam=\vptbf
     \scriptscriptfont\bffam=\vptbf
     \@fontstyleinit
     \def\rm{\viptrm\fam=\z@}%
     \def\bf{\viptbf\fam=\bffam}%
     \def\oldstyle{\viptmit\fam=\@ne}%
     \rm\fi}


\font\viiptrm=cmr7
\font\viiptmit=cmmi7
\font\viiptsy=cmsy7
\font\viiptit=cmti7
\font\viiptbf=cmbx7

\skewchar\viiptmit='177 \skewchar\viiptsy='60
\fontdimen16 \viiptsy=\the\fontdimen17 \viiptsy

\def\viipt{\ifmmode\err@badsizechange\else
     \@mathfontinit
     \textfont0=\viiptrm  \scriptfont0=\vptrm  \scriptscriptfont0=\vptrm
     \textfont1=\viiptmit \scriptfont1=\vptmit \scriptscriptfont1=\vptmit
     \textfont2=\viiptsy  \scriptfont2=\vptsy  \scriptscriptfont2=\vptsy
     \textfont3=\xptex    \scriptfont3=\xptex  \scriptscriptfont3=\xptex
     \textfont\itfam=\viiptit
     \scriptfont\itfam=\viiptit
     \scriptscriptfont\itfam=\viiptit
     \textfont\bffam=\viiptbf
     \scriptfont\bffam=\vptbf
     \scriptscriptfont\bffam=\vptbf
     \@fontstyleinit
     \def\rm{\viiptrm\fam=\z@}%
     \def\it{\viiptit\fam=\itfam}%
     \def\bf{\viiptbf\fam=\bffam}%
     \def\oldstyle{\viiptmit\fam=\@ne}%
     \rm\fi}


\font\viiiptrm=cmr8
\font\viiiptmit=cmmi8
\font\viiiptsy=cmsy8
\font\viiiptit=cmti8
\font\viiiptbf=cmbx8

\skewchar\viiiptmit='177 \skewchar\viiiptsy='60
\fontdimen16 \viiiptsy=\the\fontdimen17 \viiiptsy

\def\viiipt{\ifmmode\err@badsizechange\else
     \@mathfontinit
     \textfont0=\viiiptrm  \scriptfont0=\viptrm  \scriptscriptfont0=\vptrm
     \textfont1=\viiiptmit \scriptfont1=\viptmit \scriptscriptfont1=\vptmit
     \textfont2=\viiiptsy  \scriptfont2=\viptsy  \scriptscriptfont2=\vptsy
     \textfont3=\xptex     \scriptfont3=\xptex   \scriptscriptfont3=\xptex
     \textfont\itfam=\viiiptit
     \scriptfont\itfam=\viiptit
     \scriptscriptfont\itfam=\viiptit
     \textfont\bffam=\viiiptbf
     \scriptfont\bffam=\viptbf
     \scriptscriptfont\bffam=\vptbf
     \@fontstyleinit
     \def\rm{\viiiptrm\fam=\z@}%
     \def\it{\viiiptit\fam=\itfam}%
     \def\bf{\viiiptbf\fam=\bffam}%
     \def\oldstyle{\viiiptmit\fam=\@ne}%
     \rm\fi}


\def\getixpt{%
     \font\ixptrm=cmr9
     \font\ixptmit=cmmi9
     \font\ixptsy=cmsy9
     \font\ixptit=cmti9
     \font\ixptbf=cmbx9
     \skewchar\ixptmit='177 \skewchar\ixptsy='60
     \fontdimen16 \ixptsy=\the\fontdimen17 \ixptsy}

\def\ixpt{\ifmmode\err@badsizechange\else
     \@mathfontinit
     \textfont0=\ixptrm  \scriptfont0=\viiptrm  \scriptscriptfont0=\vptrm
     \textfont1=\ixptmit \scriptfont1=\viiptmit \scriptscriptfont1=\vptmit
     \textfont2=\ixptsy  \scriptfont2=\viiptsy  \scriptscriptfont2=\vptsy
     \textfont3=\xptex   \scriptfont3=\xptex    \scriptscriptfont3=\xptex
     \textfont\itfam=\ixptit
     \scriptfont\itfam=\viiptit
     \scriptscriptfont\itfam=\viiptit
     \textfont\bffam=\ixptbf
     \scriptfont\bffam=\viiptbf
     \scriptscriptfont\bffam=\vptbf
     \@fontstyleinit
     \def\rm{\ixptrm\fam=\z@}%
     \def\it{\ixptit\fam=\itfam}%
     \def\bf{\ixptbf\fam=\bffam}%
     \def\oldstyle{\ixptmit\fam=\@ne}%
     \rm\fi}


\font\xptrm=cmr10
\font\xptmit=cmmi10
\font\xptsy=cmsy10
\font\xptex=cmex10
\font\xptit=cmti10
\font\xptsl=cmsl10
\font\xptbf=cmbx10
\font\xpttt=cmtt10
\font\xptss=cmss10
\font\xptsc=cmcsc10
\font\xptbfs=cmbx10
\font\xptbmit=cmmib10

\skewchar\xptmit='177 \skewchar\xptbmit='177 \skewchar\xptsy='60
\fontdimen16 \xptsy=\the\fontdimen17 \xptsy

\def\xpt{\ifmmode\err@badsizechange\else
     \@mathfontinit
     \textfont0=\xptrm  \scriptfont0=\viiptrm  \scriptscriptfont0=\vptrm
     \textfont1=\xptmit \scriptfont1=\viiptmit \scriptscriptfont1=\vptmit
     \textfont2=\xptsy  \scriptfont2=\viiptsy  \scriptscriptfont2=\vptsy
     \textfont3=\xptex  \scriptfont3=\xptex    \scriptscriptfont3=\xptex
     \textfont\itfam=\xptit
     \scriptfont\itfam=\viiptit
     \scriptscriptfont\itfam=\viiptit
     \textfont\bffam=\xptbf
     \scriptfont\bffam=\viiptbf
     \scriptscriptfont\bffam=\vptbf
     \textfont\bfsfam=\xptbfs
     \scriptfont\bfsfam=\viiptbf
     \scriptscriptfont\bfsfam=\vptbf
     \textfont\bmitfam=\xptbmit
     \scriptfont\bmitfam=\viiptmit
     \scriptscriptfont\bmitfam=\vptmit
     \@fontstyleinit
     \def\rm{\xptrm\fam=\z@}%
     \def\it{\xptit\fam=\itfam}%
     \def\sl{\xptsl}%
     \def\bf{\xptbf\fam=\bffam}%
     \def\tt{\xpttt}%
     \def\ss{\xptss}%
     \def\sc{\xptsc}%
     \def\bfs{\xptbfs\fam=\bfsfam}%
     \def\bmit{\fam=\bmitfam}%
     \def\oldstyle{\xptmit\fam=\@ne}%
     \rm\fi}


\def\getxipt{%
     \font\xiptrm=cmr10  scaled\magstephalf
     \font\xiptmit=cmmi10 scaled\magstephalf
     \font\xiptsy=cmsy10 scaled\magstephalf
     \font\xiptex=cmex10 scaled\magstephalf
     \font\xiptit=cmti10 scaled\magstephalf
     \font\xiptsl=cmsl10 scaled\magstephalf
     \font\xiptbf=cmbx10 scaled\magstephalf
     \font\xipttt=cmtt10 scaled\magstephalf
     \font\xiptss=cmss10 scaled\magstephalf
     \skewchar\xiptmit='177 \skewchar\xiptsy='60
     \fontdimen16 \xiptsy=\the\fontdimen17 \xiptsy}

\def\xipt{\ifmmode\err@badsizechange\else
     \@mathfontinit
     \textfont0=\xiptrm  \scriptfont0=\viiiptrm  \scriptscriptfont0=\viptrm
     \textfont1=\xiptmit \scriptfont1=\viiiptmit \scriptscriptfont1=\viptmit
     \textfont2=\xiptsy  \scriptfont2=\viiiptsy  \scriptscriptfont2=\viptsy
     \textfont3=\xiptex  \scriptfont3=\xptex     \scriptscriptfont3=\xptex
     \textfont\itfam=\xiptit
     \scriptfont\itfam=\viiiptit
     \scriptscriptfont\itfam=\viiptit
     \textfont\bffam=\xiptbf
     \scriptfont\bffam=\viiiptbf
     \scriptscriptfont\bffam=\viptbf
     \@fontstyleinit
     \def\rm{\xiptrm\fam=\z@}%
     \def\it{\xiptit\fam=\itfam}%
     \def\sl{\xiptsl}%
     \def\bf{\xiptbf\fam=\bffam}%
     \def\tt{\xipttt}%
     \def\ss{\xiptss}%
     \def\oldstyle{\xiptmit\fam=\@ne}%
     \rm\fi}


\font\xiiptrm=cmr12
\font\xiiptmit=cmmi12
\font\xiiptsy=cmsy10  scaled\magstep1
\font\xiiptex=cmex10  scaled\magstep1
\font\xiiptit=cmti12
\font\xiiptsl=cmsl12
\font\xiiptbf=cmbx12
\font\xiipttt=cmtt12
\font\xiiptss=cmss12
\font\xiiptsc=cmcsc10 scaled\magstep1
\font\xiiptbfs=cmbx10  scaled\magstep1
\font\xiiptbmit=cmmib10 scaled\magstep1

\skewchar\xiiptmit='177 \skewchar\xiiptbmit='177 \skewchar\xiiptsy='60
\fontdimen16 \xiiptsy=\the\fontdimen17 \xiiptsy

\def\xiipt{\ifmmode\err@badsizechange\else
     \@mathfontinit
     \textfont0=\xiiptrm  \scriptfont0=\viiiptrm  \scriptscriptfont0=\viptrm
     \textfont1=\xiiptmit \scriptfont1=\viiiptmit \scriptscriptfont1=\viptmit
     \textfont2=\xiiptsy  \scriptfont2=\viiiptsy  \scriptscriptfont2=\viptsy
     \textfont3=\xiiptex  \scriptfont3=\xptex     \scriptscriptfont3=\xptex
     \textfont\itfam=\xiiptit
     \scriptfont\itfam=\viiiptit
     \scriptscriptfont\itfam=\viiptit
     \textfont\bffam=\xiiptbf
     \scriptfont\bffam=\viiiptbf
     \scriptscriptfont\bffam=\viptbf
     \textfont\bfsfam=\xiiptbfs
     \scriptfont\bfsfam=\viiiptbf
     \scriptscriptfont\bfsfam=\viptbf
     \textfont\bmitfam=\xiiptbmit
     \scriptfont\bmitfam=\viiiptmit
     \scriptscriptfont\bmitfam=\viptmit
     \@fontstyleinit
     \def\rm{\xiiptrm\fam=\z@}%
     \def\it{\xiiptit\fam=\itfam}%
     \def\sl{\xiiptsl}%
     \def\bf{\xiiptbf\fam=\bffam}%
     \def\tt{\xiipttt}%
     \def\ss{\xiiptss}%
     \def\sc{\xiiptsc}%
     \def\bfs{\xiiptbfs\fam=\bfsfam}%
     \def\bmit{\fam=\bmitfam}%
     \def\oldstyle{\xiiptmit\fam=\@ne}%
     \rm\fi}


\def\getxiiipt{%
     \font\xiiiptrm=cmr12  scaled\magstephalf
     \font\xiiiptmit=cmmi12 scaled\magstephalf
     \font\xiiiptsy=cmsy9  scaled\magstep2
     \font\xiiiptit=cmti12 scaled\magstephalf
     \font\xiiiptsl=cmsl12 scaled\magstephalf
     \font\xiiiptbf=cmbx12 scaled\magstephalf
     \font\xiiipttt=cmtt12 scaled\magstephalf
     \font\xiiiptss=cmss12 scaled\magstephalf
     \skewchar\xiiiptmit='177 \skewchar\xiiiptsy='60
     \fontdimen16 \xiiiptsy=\the\fontdimen17 \xiiiptsy}

\def\xiiipt{\ifmmode\err@badsizechange\else
     \@mathfontinit
     \textfont0=\xiiiptrm  \scriptfont0=\xptrm  \scriptscriptfont0=\viiptrm
     \textfont1=\xiiiptmit \scriptfont1=\xptmit \scriptscriptfont1=\viiptmit
     \textfont2=\xiiiptsy  \scriptfont2=\xptsy  \scriptscriptfont2=\viiptsy
     \textfont3=\xivptex   \scriptfont3=\xptex  \scriptscriptfont3=\xptex
     \textfont\itfam=\xiiiptit
     \scriptfont\itfam=\xptit
     \scriptscriptfont\itfam=\viiptit
     \textfont\bffam=\xiiiptbf
     \scriptfont\bffam=\xptbf
     \scriptscriptfont\bffam=\viiptbf
     \@fontstyleinit
     \def\rm{\xiiiptrm\fam=\z@}%
     \def\it{\xiiiptit\fam=\itfam}%
     \def\sl{\xiiiptsl}%
     \def\bf{\xiiiptbf\fam=\bffam}%
     \def\tt{\xiiipttt}%
     \def\ss{\xiiiptss}%
     \def\oldstyle{\xiiiptmit\fam=\@ne}%
     \rm\fi}


\font\xivptrm=cmr12   scaled\magstep1
\font\xivptmit=cmmi12  scaled\magstep1
\font\xivptsy=cmsy10  scaled\magstep2
\font\xivptex=cmex10  scaled\magstep2
\font\xivptit=cmti12  scaled\magstep1
\font\xivptsl=cmsl12  scaled\magstep1
\font\xivptbf=cmbx12  scaled\magstep1
\font\xivpttt=cmtt12  scaled\magstep1
\font\xivptss=cmss12  scaled\magstep1
\font\xivptsc=cmcsc10 scaled\magstep2
\font\xivptbfs=cmbx10  scaled\magstep2
\font\xivptbmit=cmmib10 scaled\magstep2

\skewchar\xivptmit='177 \skewchar\xivptbmit='177 \skewchar\xivptsy='60
\fontdimen16 \xivptsy=\the\fontdimen17 \xivptsy

\def\xivpt{\ifmmode\err@badsizechange\else
     \@mathfontinit
     \textfont0=\xivptrm  \scriptfont0=\xptrm  \scriptscriptfont0=\viiptrm
     \textfont1=\xivptmit \scriptfont1=\xptmit \scriptscriptfont1=\viiptmit
     \textfont2=\xivptsy  \scriptfont2=\xptsy  \scriptscriptfont2=\viiptsy
     \textfont3=\xivptex  \scriptfont3=\xptex  \scriptscriptfont3=\xptex
     \textfont\itfam=\xivptit
     \scriptfont\itfam=\xptit
     \scriptscriptfont\itfam=\viiptit
     \textfont\bffam=\xivptbf
     \scriptfont\bffam=\xptbf
     \scriptscriptfont\bffam=\viiptbf
     \textfont\bfsfam=\xivptbfs
     \scriptfont\bfsfam=\xptbfs
     \scriptscriptfont\bfsfam=\viiptbf
     \textfont\bmitfam=\xivptbmit
     \scriptfont\bmitfam=\xptbmit
     \scriptscriptfont\bmitfam=\viiptmit
     \@fontstyleinit
     \def\rm{\xivptrm\fam=\z@}%
     \def\it{\xivptit\fam=\itfam}%
     \def\sl{\xivptsl}%
     \def\bf{\xivptbf\fam=\bffam}%
     \def\tt{\xivpttt}%
     \def\ss{\xivptss}%
     \def\sc{\xivptsc}%
     \def\bfs{\xivptbfs\fam=\bfsfam}%
     \def\bmit{\fam=\bmitfam}%
     \def\oldstyle{\xivptmit\fam=\@ne}%
     \rm\fi}


\font\xviiptrm=cmr17
\font\xviiptmit=cmmi12 scaled\magstep2
\font\xviiptsy=cmsy10 scaled\magstep3
\font\xviiptex=cmex10 scaled\magstep3
\font\xviiptit=cmti12 scaled\magstep2
\font\xviiptbf=cmbx12 scaled\magstep2
\font\xviiptbfs=cmbx10 scaled\magstep3

\skewchar\xviiptmit='177 \skewchar\xviiptsy='60
\fontdimen16 \xviiptsy=\the\fontdimen17 \xviiptsy

\def\xviipt{\ifmmode\err@badsizechange\else
     \@mathfontinit
     \textfont0=\xviiptrm  \scriptfont0=\xiiptrm  \scriptscriptfont0=\viiiptrm
     \textfont1=\xviiptmit \scriptfont1=\xiiptmit \scriptscriptfont1=\viiiptmit
     \textfont2=\xviiptsy  \scriptfont2=\xiiptsy  \scriptscriptfont2=\viiiptsy
     \textfont3=\xviiptex  \scriptfont3=\xiiptex  \scriptscriptfont3=\xptex
     \textfont\itfam=\xviiptit
     \scriptfont\itfam=\xiiptit
     \scriptscriptfont\itfam=\viiiptit
     \textfont\bffam=\xviiptbf
     \scriptfont\bffam=\xiiptbf
     \scriptscriptfont\bffam=\viiiptbf
     \textfont\bfsfam=\xviiptbfs
     \scriptfont\bfsfam=\xiiptbfs
     \scriptscriptfont\bfsfam=\viiiptbf
     \@fontstyleinit
     \def\rm{\xviiptrm\fam=\z@}%
     \def\it{\xviiptit\fam=\itfam}%
     \def\bf{\xviiptbf\fam=\bffam}%
     \def\bfs{\xviiptbfs\fam=\bfsfam}%
     \def\oldstyle{\xviiptmit\fam=\@ne}%
     \rm\fi}


\font\xxiptrm=cmr17  scaled\magstep1


\def\xxipt{\ifmmode\err@badsizechange\else
     \@mathfontinit
     \@fontstyleinit
     \def\rm{\xxiptrm\fam=\z@}%
     \rm\fi}


\font\xxvptrm=cmr17  scaled\magstep2


\def\xxvpt{\ifmmode\err@badsizechange\else
     \@mathfontinit
     \@fontstyleinit
     \def\rm{\xxvptrm\fam=\z@}%
     \rm\fi}




\message{Loading jyTeX macros...}

\message{modifications to plain.tex,}


\def\newcount{\alloc@0\count\countdef\insc@unt}
\def\newdimen{\alloc@1\dimen\dimendef\insc@unt}
\def\newskip{\alloc@2\skip\skipdef\insc@unt}
\def\newmuskip{\alloc@3\muskip\muskipdef\@cclvi}
\def\newbox{\alloc@4\box\chardef\insc@unt}
\def\newtoks{\alloc@5\toks\toksdef\@cclvi}
\def\newhelp#1#2{\newtoks#1\global#1\expandafter{\csname#2\endcsname}}
\def\newread{\alloc@6\read\chardef\sixt@@n}
\def\newwrite{\alloc@7\write\chardef\sixt@@n}
\def\newfam{\alloc@8\fam\chardef\sixt@@n}
\def\newinsert#1{\global\advance\insc@unt by\m@ne
     \ch@ck0\insc@unt\count
     \ch@ck1\insc@unt\dimen
     \ch@ck2\insc@unt\skip
     \ch@ck4\insc@unt\box
     \allocationnumber=\insc@unt
     \global\chardef#1=\allocationnumber
     \wlog{\string#1=\string\insert\the\allocationnumber}}
\def\newif#1{\count@\escapechar \escapechar\m@ne
     \expandafter\expandafter\expandafter
          \xdef\@if#1{true}{\let\noexpand#1=\noexpand\iftrue}%
     \expandafter\expandafter\expandafter
          \xdef\@if#1{false}{\let\noexpand#1=\noexpand\iffalse}%
     \global\@if#1{false}\escapechar=\count@}


\newlinechar=`\^^J
\overfullrule=0pt




\let\itfam=\undefined

\let\bffam=\undefined

\count18=3


\chardef\sharps="19


\mathchardef\alpha="710B
\mathchardef\beta="710C
\mathchardef\gamma="710D
\mathchardef\delta="710E
\mathchardef\epsilon="710F
\mathchardef\zeta="7110
\mathchardef\eta="7111
\mathchardef\theta="7112
\mathchardef\iota="7113
\mathchardef\kappa="7114
\mathchardef\lambda="7115
\mathchardef\mu="7116
\mathchardef\nu="7117
\mathchardef\xi="7118
\mathchardef\pi="7119
\mathchardef\rho="711A
\mathchardef\sigma="711B
\mathchardef\tau="711C
\mathchardef\upsilon="711D
\mathchardef\phi="711E
\mathchardef\chi="711F
\mathchardef\psi="7120
\mathchardef\omega="7121
\mathchardef\varepsilon="7122
\mathchardef\vartheta="7123
\mathchardef\varpi="7124
\mathchardef\varrho="7125
\mathchardef\varsigma="7126
\mathchardef\varphi="7127
\mathchardef\imath="717B
\mathchardef\jmath="717C
\mathchardef\ell="7160
\mathchardef\wp="717D
\mathchardef\partial="7140
\mathchardef\flat="715B
\mathchardef\natural="715C
\mathchardef\sharp="715D



\def\angle{{\vbox{\ialign{$\m@th\scriptstyle##$\crcr
     \not\mathrel{\mkern14mu}\crcr
     \noalign{\nointerlineskip}
     \mkern2.5mu\leaders\hrule height.34\rp@\hfill\mkern2.5mu\crcr}}}}
\def\vdots{\vbox{\baselineskip4\rp@ \lineskiplimit\z@
     \kern6\rp@\hbox{.}\hbox{.}\hbox{.}}}
\def\ddots{\mathinner{\mkern1mu\raise7\rp@\vbox{\kern7\rp@\hbox{.}}\mkern2mu
     \raise4\rp@\hbox{.}\mkern2mu\raise\rp@\hbox{.}\mkern1mu}}
\def\overrightarrow#1{\vbox{\ialign{##\crcr
     \rightarrowfill\crcr
     \noalign{\kern-\rp@\nointerlineskip}
     $\hfil\displaystyle{#1}\hfil$\crcr}}}
\def\overleftarrow#1{\vbox{\ialign{##\crcr
     \leftarrowfill\crcr
     \noalign{\kern-\rp@\nointerlineskip}
     $\hfil\displaystyle{#1}\hfil$\crcr}}}
\def\overbrace#1{\mathop{\vbox{\ialign{##\crcr
     \noalign{\kern3\rp@}
     \downbracefill\crcr
     \noalign{\kern3\rp@\nointerlineskip}
     $\hfil\displaystyle{#1}\hfil$\crcr}}}\limits}
\def\underbrace#1{\mathop{\vtop{\ialign{##\crcr
     $\hfil\displaystyle{#1}\hfil$\crcr
     \noalign{\kern3\rp@\nointerlineskip}
     \upbracefill\crcr
     \noalign{\kern3\rp@}}}}\limits}
\def\big#1{{\hbox{$\left#1\vbox to8.5\rp@ {}\right.\n@space$}}}
\def\Big#1{{\hbox{$\left#1\vbox to11.5\rp@ {}\right.\n@space$}}}
\def\bigg#1{{\hbox{$\left#1\vbox to14.5\rp@ {}\right.\n@space$}}}
\def\Bigg#1{{\hbox{$\left#1\vbox to17.5\rp@ {}\right.\n@space$}}}
\def\@vereq#1#2{\lower.5\rp@\vbox{\baselineskip\z@skip\lineskip-.5\rp@
     \ialign{$\m@th#1\hfil##\hfil$\crcr#2\crcr=\crcr}}}
\def\rlh@#1{\vcenter{\hbox{\ooalign{\raise2\rp@
     \hbox{$#1\rightharpoonup$}\crcr
     $#1\leftharpoondown$}}}}
\def\bordermatrix#1{\begingroup\m@th
     \setbox\z@\vbox{%
          \def\cr{\crcr\noalign{\kern2\rp@\global\let\cr\endline}}%
          \ialign{$##$\hfil\kern2\rp@\kern\p@renwd
               &\thinspace\hfil$##$\hfil&&\quad\hfil$##$\hfil\crcr
               \omit\strut\hfil\crcr
               \noalign{\kern-\baselineskip}%
               #1\crcr\omit\strut\cr}}%
     \setbox\tw@\vbox{\unvcopy\z@\global\setbox\@ne\lastbox}%
     \setbox\tw@\hbox{\unhbox\@ne\unskip\global\setbox\@ne\lastbox}%
     \setbox\tw@\hbox{$\kern\wd\@ne\kern-\p@renwd\left(\kern-\wd\@ne
          \global\setbox\@ne\vbox{\box\@ne\kern2\rp@}%
          \vcenter{\kern-\ht\@ne\unvbox\z@\kern-\baselineskip}%
          \,\right)$}%
     \null\;\vbox{\kern\ht\@ne\box\tw@}\endgroup}
\def\endinsert{\egroup
     \if@mid\dimen@\ht\z@
          \advance\dimen@\dp\z@
          \advance\dimen@12\rp@
          \advance\dimen@\pagetotal
          \ifdim\dimen@>\pagegoal\@midfalse\p@gefalse\fi
     \fi
     \if@mid\bigskip\box\z@
          \bigbreak
     \else\insert\topins{\penalty100 \splittopskip\z@skip
               \splitmaxdepth\maxdimen\floatingpenalty\z@
               \ifp@ge\dimen@\dp\z@
                    \vbox to\vsize{\unvbox\z@\kern-\dimen@}%
               \else\box\z@\nobreak\bigskip
               \fi}%
     \fi
     \endgroup}


\def\cases#1{\left\{\,\vcenter{\m@th
     \ialign{$##\hfil$&\quad##\hfil\crcr#1\crcr}}\right.}
\def\matrix#1{\null\,\vcenter{\m@th
     \ialign{\hfil$##$\hfil&&\quad\hfil$##$\hfil\crcr
          \mathstrut\crcr
          \noalign{\kern-\baselineskip}
          #1\crcr
          \mathstrut\crcr
          \noalign{\kern-\baselineskip}}}\,}


\newif\ifraggedbottom

\def\raggedbottom{\ifraggedbottom\else
     \advance\topskip by\z@ plus60pt \raggedbottomtrue\fi}%
\def\normalbottom{\ifraggedbottom
     \advance\topskip by\z@ plus-60pt \raggedbottomfalse\fi}

\message{hacks,}


\toksdef\toks@i=1
\toksdef\toks@ii=2


\def\TeX{T\kern-.1667em \lower.5ex \hbox{E}\kern-.125em X\null}
\def\jyTeX{{\leavevmode
     \raise.587ex \hbox{\it\j}\kern-.1em \lower.048ex \hbox{\it y}\kern-.12em
     \TeX}}

\let\then=\iftrue
\def\ifnoarg#1\then{\def\hack@{#1}\ifx\hack@\empty}
\def\ifundefined#1\then{%
     \expandafter\ifx\csname\expandafter\blank\string#1\endcsname\relax}
\def\useif#1\then{\csname#1\endcsname}
\def\usename#1{\csname#1\endcsname}
\def\useafter#1#2{\expandafter#1\csname#2\endcsname}

\long\def\loop#1\repeat{\def\@iterate{#1\expandafter\@iterate\fi}\@iterate
     \let\@iterate=\relax}

\let\TeXend=\end
\def\begin#1{\begingroup\def\@@blockname{#1}\usename{begin#1}}
\def\end#1{\usename{end#1}\def\hack@{#1}%
     \ifx\@@blockname\hack@
          \endgroup
     \else\err@badgroup\hack@\@@blockname
     \fi}
\def\@@blockname{}

\def\defaultoption[#1]#2{%
     \def\hack@{\ifx\hack@ii[\toks@={#2}\else\toks@={#2[#1]}\fi\the\toks@}%
     \futurelet\hack@ii\hack@}

\def\markup#1{\let\@@marksf=\empty
     \ifhmode\edef\@@marksf{\spacefactor=\the\spacefactor\relax}\/\fi
     ${}^{\hbox{\subscriptfonts#1}}$\@@marksf}


\newtoks\shortyear
\newtoks\militaryhour
\newtoks\standardhour
\newtoks\minute
\newtoks\amorpm

\def\settime{\count@=\time\divide\count@ by60
     \militaryhour=\expandafter{\number\count@}%
     {\multiply\count@ by-60 \advance\count@ by\time
          \xdef\hack@{\ifnum\count@<10 0\fi\number\count@}}%
     \minute=\expandafter{\hack@}%
     \ifnum\count@<12
          \amorpm={am}
     \else\amorpm={pm}
          \ifnum\count@>12 \advance\count@ by-12 \fi
     \fi
     \standardhour=\expandafter{\number\count@}%
     \def\hack@19##1##2{\shortyear={##1##2}}%
          \expandafter\hack@\the\year}

\def\monthword#1{%
     \ifcase#1
          $\bullet$\err@badcountervalue{monthword}%
          \or January\or February\or March\or April\or May\or June%
          \or July\or August\or September\or October\or November\or December%
     \else$\bullet$\err@badcountervalue{monthword}%
     \fi}

\def\monthabbr#1{%
     \ifcase#1
          $\bullet$\err@badcountervalue{monthabbr}%
          \or Jan\or Feb\or Mar\or Apr\or May\or Jun%
          \or Jul\or Aug\or Sep\or Oct\or Nov\or Dec%
     \else$\bullet$\err@badcountervalue{monthabbr}%
     \fi}

\def\militarytime{\the\militaryhour:\the\minute}
\def\standardtime{\the\standardhour:\the\minute}


\def\@setnumstyle#1#2{\expandafter\global\expandafter\expandafter
     \expandafter\let\expandafter\expandafter
     \csname @\expandafter\blank\string#1style\endcsname
     \csname#2\endcsname}
\def\numstyle#1{\usename{@\expandafter\blank\string#1style}#1}
\def\ifblank#1\then{\useafter\ifx{@\expandafter\blank\string#1}\blank}

\def\blank#1{}

\def\Roman#1{\expandafter\uppercase\expandafter{\romannumeral#1}}
\def\alphabetic#1{%
     \ifcase#1
          $\bullet$\err@badcountervalue{alphabetic}%
          \or a\or b\or c\or d\or e\or f\or g\or h\or i\or j\or k\or l\or m%
          \or n\or o\or p\or q\or r\or s\or t\or u\or v\or w\or x\or y\or z%
     \else$\bullet$\err@badcountervalue{alphabetic}%
     \fi}
\def\Alphabetic#1{\expandafter\uppercase\expandafter{\alphabetic{#1}}}
\def\symbols#1{%
     \ifcase#1
          $\bullet$\err@badcountervalue{symbols}%
          \or*\or\dag\or\ddag\or\S\or$\|$%
          \or**\or\dag\dag\or\ddag\ddag\or\S\S\or$\|\|$%
     \else$\bullet$\err@badcountervalue{symbols}%
     \fi}


\catcode`\^^?=13 \def^^?{\relax}

\def\trimleading#1\to#2{\edef#2{#1}%
     \expandafter\@trimleading\expandafter#2#2^^?^^?}
\def\@trimleading#1#2#3^^?{\ifx#2^^?\def#1{}\else\def#1{#2#3}\fi}

\def\trimtrailing#1\to#2{\edef#2{#1}%
     \expandafter\@trimtrailing\expandafter#2#2^^? ^^?\relax}
\def\@trimtrailing#1#2 ^^?#3{\ifx#3\relax\toks@={}%
     \else\def#1{#2}\toks@={\trimtrailing#1\to#1}\fi
     \the\toks@}

\def\trim#1\to#2{\trimleading#1\to#2\trimtrailing#2\to#2}

\catcode`\^^?=15


\long\def\additemL#1\to#2{\toks@={\^^\{#1}}\toks@ii=\expandafter{#2}%
     \xdef#2{\the\toks@\the\toks@ii}}

\long\def\additemR#1\to#2{\toks@={\^^\{#1}}\toks@ii=\expandafter{#2}%
     \xdef#2{\the\toks@ii\the\toks@}}

\def\getitemL#1\to#2{\expandafter\@getitemL#1\hack@#1#2}
\def\@getitemL\^^\#1#2\hack@#3#4{\def#4{#1}\def#3{#2}}

\message{font macros,}


\newdimen\rp@
\newcount\@@sizeindex \@@sizeindex=0
\newcount\@@factori
\newcount\@@factorii
\newcount\@@factoriii
\newcount\@@factoriv

\countdef\maxfam=18
\newfam\itfam
\newfam\bffam
\newfam\bfsfam
\newfam\bmitfam

\def\@mathfontinit{\count@=4
     \loop\textfont\count@=\nullfont
          \scriptfont\count@=\nullfont
          \scriptscriptfont\count@=\nullfont
          \ifnum\count@<\maxfam\advance\count@ by\@ne
     \repeat}

\def\@fontstyleinit{%
     \def\it{\err@fontnotavailable\it}%
     \def\bf{\err@fontnotavailable\bf}%
     \def\bfs{\err@bfstobf}%
     \def\bmit{\err@fontnotavailable\bmit}%
     \def\sc{\err@fontnotavailable\sc}%
     \def\sl{\err@sltoit}%
     \def\ss{\err@fontnotavailable\ss}%
     \def\tt{\err@fontnotavailable\tt}}

\def\@parameterinit#1{\rm\rp@=.1em \@getscaling{#1}%
     \let\^^\=\@doscaling\scalingskipslist
     \setbox\strutbox=\hbox{\vrule
          height.708\baselineskip depth.292\baselineskip width\z@}}

\def\@getfactor#1#2#3#4{\@@factori=#1 \@@factorii=#2
     \@@factoriii=#3 \@@factoriv=#4}

\def\@getscaling#1{\count@=#1 \advance\count@ by-\@@sizeindex\@@sizeindex=#1
     \ifnum\count@<0
          \let\@mulordiv=\divide
          \let\@divormul=\multiply
          \multiply\count@ by\m@ne
     \else\let\@mulordiv=\multiply
          \let\@divormul=\divide
     \fi
     \edef\@@scratcha{\ifcase\count@                {1}{1}{1}{1}\or
          {1}{7}{23}{3}\or     {2}{5}{3}{1}\or      {9}{89}{13}{1}\or
          {6}{25}{6}{1}\or     {8}{71}{14}{1}\or    {6}{25}{36}{5}\or
          {1}{7}{53}{4}\or     {12}{125}{108}{5}\or {3}{14}{53}{5}\or
          {6}{41}{17}{1}\or    {13}{31}{13}{2}\or   {9}{107}{71}{2}\or
          {11}{139}{124}{3}\or {1}{6}{43}{2}\or     {10}{107}{42}{1}\or
          {1}{5}{43}{2}\or     {5}{69}{65}{1}\or    {11}{97}{91}{2}\fi}%
     \expandafter\@getfactor\@@scratcha}

\def\@doscaling#1{\@mulordiv#1by\@@factori\@divormul#1by\@@factorii
     \@mulordiv#1by\@@factoriii\@divormul#1by\@@factoriv}


\newskip\headskip
\newskip\footskip

\def\typesize=#1pt{\count@=#1 \advance\count@ by-10
     \ifcase\count@
          \@setsizex\or\err@badtypesize\or
          \@setsizexii\or\err@badtypesize\or
          \@setsizexiv
     \else\err@badtypesize
     \fi}

\def\@setsizex{\getixpt
     \def\subsubscriptfonts{\vpt}%
          \def\subsubscriptsize{\vpt\@parameterinit{-8}}%
     \def\subscriptfonts{\viipt}\def\subscriptsize{\viipt\@parameterinit{-4}}%
     \def\footnotefonts{\viiipt}\def\footnotesize{\viiipt\@parameterinit{-2}}%
     \def\smallfonts{\ixpt}\def\smallsize{\ixpt\@parameterinit{-1}}%
     \def\normalfonts{\xpt}\def\normalsize{\xpt\@parameterinit{0}}%
     \def\bigfonts{\xiipt}\def\bigsize{\xiipt\@parameterinit{2}}%
     \def\Bigfonts{\xivpt}\def\Bigsize{\xivpt\@parameterinit{4}}%
     \def\biggfonts{\xviipt}\def\biggsize{\xviipt\@parameterinit{6}}%
     \def\Biggfonts{\xxipt}\def\Biggsize{\xxipt\@parameterinit{8}}%
     \def\tinyfonts{\vpt}\def\tinysize{\vpt\@parameterinit{-8}}%
     \def\HUGEFONTS{\xxvpt}\def\HUGESIZE{\xxvpt\@parameterinit{10}}%
     \normalsize\fixedskipslist}

\def\@setsizexii{\getxipt
     \def\subsubscriptfonts{\vipt}%
          \def\subsubscriptsize{\vipt\@parameterinit{-6}}%
     \def\subscriptfonts{\viiipt}%
          \def\subscriptsize{\viiipt\@parameterinit{-2}}%
     \def\footnotefonts{\xpt}\def\footnotesize{\xpt\@parameterinit{0}}%
     \def\smallfonts{\xipt}\def\smallsize{\xipt\@parameterinit{1}}%
     \def\normalfonts{\xiipt}\def\normalsize{\xiipt\@parameterinit{2}}%
     \def\bigfonts{\xivpt}\def\bigsize{\xivpt\@parameterinit{4}}%
     \def\Bigfonts{\xviipt}\def\Bigsize{\xviipt\@parameterinit{6}}%
     \def\biggfonts{\xxipt}\def\biggsize{\xxipt\@parameterinit{8}}%
     \def\Biggfonts{\xxvpt}\def\Biggsize{\xxvpt\@parameterinit{10}}%
     \def\tinyfonts{\vpt}\def\tinysize{\vpt\@parameterinit{-8}}%
     \def\HUGEFONTS{\xxvpt}\def\HUGESIZE{\xxvpt\@parameterinit{10}}%
     \normalsize\fixedskipslist}

\def\@setsizexiv{\getxiiipt
     \def\subsubscriptfonts{\viipt}%
          \def\subsubscriptsize{\viipt\@parameterinit{-4}}%
     \def\subscriptfonts{\xpt}\def\subscriptsize{\xpt\@parameterinit{0}}%
     \def\footnotefonts{\xiipt}\def\footnotesize{\xiipt\@parameterinit{2}}%
     \def\smallfonts{\xiiipt}\def\smallsize{\xiiipt\@parameterinit{3}}%
     \def\normalfonts{\xivpt}\def\normalsize{\xivpt\@parameterinit{4}}%
     \def\bigfonts{\xviipt}\def\bigsize{\xviipt\@parameterinit{6}}%
     \def\Bigfonts{\xxipt}\def\Bigsize{\xxipt\@parameterinit{8}}%
     \def\biggfonts{\xxvpt}\def\biggsize{\xxvpt\@parameterinit{10}}%
     \def\Biggfonts{\err@sizetoolarge\Biggfonts\HUGEFONTS}%
          \def\Biggsize{\err@sizetoolarge\Biggsize\HUGESIZE}%
     \def\tinyfonts{\vpt}\def\tinysize{\vpt\@parameterinit{-8}}%
     \def\HUGEFONTS{\xxvpt}\def\HUGESIZE{\xxvpt\@parameterinit{10}}%
     \normalsize\fixedskipslist}

\def\subsubscriptfonts{\vpt} \def\subsubscriptsize{\vpt\@parameterinit{-8}}
\def\subscriptfonts{\viipt}  \def\subscriptsize{\viipt\@parameterinit{-4}}
\def\footnotefonts{\viiipt}  \def\footnotesize{\viiipt\@parameterinit{-2}}
\def\smallfonts{\err@sizenotavailable\smallfonts}
                             \def\smallsize{\ixpt\@parameterinit{-1}}
\def\normalfonts{\xpt}       \def\normalsize{\xpt\@parameterinit{0}}
\def\bigfonts{\xiipt}        \def\bigsize{\xiipt\@parameterinit{2}}
\def\Bigfonts{\xivpt}        \def\Bigsize{\xivpt\@parameterinit{4}}
\def\biggfonts{\xviipt}      \def\biggsize{\xviipt\@parameterinit{6}}
\def\Biggfonts{\xxipt}       \def\Biggsize{\xxipt\@parameterinit{8}}
\def\tinyfonts{\vpt}         \def\tinysize{\vpt\@parameterinit{-8}}
\def\HUGEFONTS{\xxvpt}       \def\HUGESIZE{\xxvpt\@parameterinit{10}}

\message{document layout,}


\newtoks\everyoutput \everyoutput={}
\newdimen\depthofpage
\newcount\pagenum \pagenum=0

\newdimen\oddtopmargin  \newdimen\eventopmargin
\newdimen\oddleftmargin \newdimen\evenleftmargin
\newtoks\oddhead        \newtoks\evenhead
\newtoks\oddfoot        \newtoks\evenfoot

\def\topmargin{\afterassignment\@seteventop\oddtopmargin}
\def\leftmargin{\afterassignment\@setevenleft\oddleftmargin}
\def\head{\afterassignment\@setevenhead\oddhead}
\def\foot{\afterassignment\@setevenfoot\oddfoot}

\def\@seteventop{\eventopmargin=\oddtopmargin}
\def\@setevenleft{\evenleftmargin=\oddleftmargin}
\def\@setevenhead{\evenhead=\oddhead}
\def\@setevenfoot{\evenfoot=\oddfoot}

\def\pagenumstyle#1{\@setnumstyle\pagenum{#1}}

\newif\ifdraft
\def\draft{\drafttrue\leftmargin=.5in \overfullrule=5pt }

\def\outputstyle#1{\global\expandafter\let\expandafter
          \@outputstyle\csname#1output\endcsname
     \usename{#1setup}}

\output={\@outputstyle}

\def\normaloutput{\the\everyoutput
     \global\advance\pagenum by\@ne
     \ifodd\pagenum
          \voffset=\oddtopmargin \hoffset=\oddleftmargin
     \else\voffset=\eventopmargin \hoffset=\evenleftmargin
     \fi
     \advance\voffset by-1in  \advance\hoffset by-1in
     \count0=\pagenum
     \expandafter\shipout\pagebox
     \ifnum\outputpenalty>-\@MM\else\dosupereject\fi}

\newdimen\fullhsize
\newbox\leftpage
\newcount\leftpagenum
\newcount\outputpagenum \outputpagenum=0
\let\leftorright=L

\def\twoupoutput{\the\everyoutput
     \global\advance\pagenum by\@ne
     \if L\leftorright
          \global\setbox\leftpage=\leftline{\pagebox}%
          \global\leftpagenum=\pagenum
          \global\let\leftorright=R%
     \else\global\advance\outputpagenum by\@ne
          \ifodd\outputpagenum
               \voffset=\oddtopmargin \hoffset=\oddleftmargin
          \else\voffset=\eventopmargin \hoffset=\evenleftmargin
          \fi
          \advance\voffset by-1in  \advance\hoffset by-1in
          \count0=\leftpagenum \count1=\pagenum
          \shipout\vbox{\hbox to\fullhsize
               {\box\leftpage\hfil\leftline{\pagebox}}}%
          \global\let\leftorright=L%
     \fi
     \ifnum\outputpenalty>-\@MM
     \else\dosupereject
          \if R\leftorright
               \globaldefs=\@ne\head={\hfil}\foot={\hfil}\globaldefs=\z@
               \null\newpage
          \fi
     \fi}

\def\pagebox{\vbox{\makeheadline\pagebody\makefootline}}

\def\makeheadline{%
     \vbox to\z@{\baselinestretch=\@m
          \vskip\topskip\vskip-.708\baselineskip\vskip-\headskip
          \line{\vbox to\ht\strutbox{}%
               \ifodd\pagenum\the\oddhead\else\the\evenhead\fi}%
          \vss}%
     \nointerlineskip}

\def\pagebody{\vbox to\vsize{%
     \boxmaxdepth\maxdepth
     \ifvoid\topins\else\unvbox\topins\fi
     \depthofpage=\dp255
     \unvbox255
     \ifraggedbottom\kern-\depthofpage\vfil\fi
     \ifvoid\footins
     \else\vskip\skip\footins
          \footnoterule
          \unvbox\footins
          \vskip-\footnoteskip
     \fi}}

\def\makefootline{\baselineskip=\footskip
     \line{\ifodd\pagenum\the\oddfoot\else\the\evenfoot\fi}}


\newskip\abovechapterskip
\newskip\belowchapterskip
\newskip\abovesectionskip
\newskip\belowsectionskip
\newskip\abovesubsectionskip
\newskip\belowsubsectionskip

\def\chapterstyle#1{\global\expandafter\let\expandafter\@chapterstyle
     \csname#1text\endcsname}
\def\sectionstyle#1{\global\expandafter\let\expandafter\@sectionstyle
     \csname#1text\endcsname}
\def\subsectionstyle#1{\global\expandafter\let\expandafter\@subsectionstyle
     \csname#1text\endcsname}

\def\chapter#1{%
     \ifdim\lastskip=17sp \else\chapterbreak\vskip\abovechapterskip\fi
     \@chapterstyle{\ifblank\chapternumstyle\then
          \else\newchapternum=\next\chapternumformat\ \fi#1}%
     \nobreak\vskip\belowchapterskip\vskip17sp }

\def\section#1{%
     \ifdim\lastskip=17sp \else\sectionbreak\vskip\abovesectionskip\fi
     \@sectionstyle{\ifblank\sectionnumstyle\then
          \else\newsectionnum=\next\sectionnumformat\ \fi#1}%
     \nobreak\vskip\belowsectionskip\vskip17sp }

\def\subsection#1{%
     \ifdim\lastskip=17sp \else\subsectionbreak\vskip\abovesubsectionskip\fi
     \@subsectionstyle{\ifblank\subsectionnumstyle\then
          \else\newsubsectionnum=\next\subsectionnumformat\ \fi#1}%
     \nobreak\vskip\belowsubsectionskip\vskip17sp }


\let\TeXunderline=\underline
\let\TeXoverline=\overline
\def\underline#1{\relax\ifmmode\TeXunderline{#1}\else
     $\TeXunderline{\hbox{#1}}$\fi}
\def\overline#1{\relax\ifmmode\TeXoverline{#1}\else
     $\TeXoverline{\hbox{#1}}$\fi}

\def\baselinestretch{\afterassignment\@baselinestretch\count@}
\def\@baselinestretch{\baselineskip=\normalbaselineskip
     \divide\baselineskip by\@m\baselineskip=\count@\baselineskip
     \setbox\strutbox=\hbox{\vrule
          height.708\baselineskip depth.292\baselineskip width\z@}%
     \bigskipamount=\the\baselineskip
          plus.25\baselineskip minus.25\baselineskip
     \medskipamount=.5\baselineskip
          plus.125\baselineskip minus.125\baselineskip
     \smallskipamount=.25\baselineskip
          plus.0625\baselineskip minus.0625\baselineskip}

\def\\{\ifhmode\ifnum\lastpenalty=-\@M\else\hfil\penalty-\@M\fi\fi
     \ignorespaces}
\def\newpage{\vfil\break}

\def\lefttext#1{\par{\@text\leftskip=\z@\rightskip=\centering
     \noindent#1\par}}
\def\righttext#1{\par{\@text\leftskip=\centering\rightskip=\z@
     \noindent#1\par}}
\def\centertext#1{\par{\@text\leftskip=\centering\rightskip=\centering
     \noindent#1\par}}
\def\@text{\parindent=\z@ \parfillskip=\z@ \everypar={}%
     \spaceskip=.3333em \xspaceskip=.5em
     \def\\{\ifhmode\ifnum\lastpenalty=-\@M\else\penalty-\@M\fi\fi
          \ignorespaces}}

\def\beginleft{\par\@text\leftskip=\z@ \rightskip=\centering}
     
\def\beginright{\par\@text\leftskip=\centering\rightskip=\z@ }
     
\def\begincenter{\par\@text\leftskip=\centering\rightskip=\centering}

\def\beginnarrow{\defaultoption[\parindent]\@beginnarrow}
\def\@beginnarrow[#1]{\par\advance\leftskip by#1\advance\rightskip by#1}

\begingroup
\catcode`\[=1 \catcode`\{=11
\gdef\beginignore[\endgroup\bgroup
     \catcode`\e=0 \catcode`\\=12 \catcode`\{=11 \catcode`\f=12 \let\or=\relax
     \let\nd{ignor=\fi \let\}=\egroup
     \iffalse}
\endgroup

\long\def\marginnote#1{\leavevmode
     \edef\@marginsf{\spacefactor=\the\spacefactor\relax}%
     \ifdraft\strut\vadjust{%
          \hbox to\z@{\hskip\hsize\hskip.1in
               \vbox to\z@{\vskip-\dp\strutbox
                    \marginnoteformat
                    \vskip-\ht\strutbox
                    \noindent\strut#1\par
                    \vss}%
               \hss}}%
     \fi
     \@marginsf}


\newtoks\everybye \everybye={\par\vfil}
\outer\def\bye{\the\everybye
     \footnotecheck
     \prelabelcheck
     \streamcheck
     \supereject
     \TeXend}

\message{footnotes,}

\newcount\footnotenum \footnotenum=0
\newskip\footnoteskip
\let\@footnotelist=\empty

\def\footnotenumstyle#1{\@setnumstyle\footnotenum{#1}%
     \useafter\ifx{@footnotenumstyle}\symbols
          \global\let\@footup=\empty
     \else\global\let\@footup=\markup
     \fi}

\def\footnote{\footnotecheck\defaultoption[]\@footnote}
\def\@footnote[#1]{\@footnotemark[#1]\@footnotetext}

\def\footnotemark{\defaultoption[]\@footnotemark}
\def\@footnotemark[#1]{\let\@footsf=\empty
     \ifhmode\edef\@footsf{\spacefactor=\the\spacefactor\relax}\/\fi
     \ifnoarg#1\then
          \global\advance\footnotenum by\@ne
          \@footup{\footnotenumformat}%
          \edef\@@foota{\footnotenum=\the\footnotenum\relax}%
          \expandafter\additemR\expandafter\@footup\expandafter
               {\@@foota\footnotenumformat}\to\@footnotelist
          \global\let\@footnotelist=\@footnotelist
     \else\markup{#1}%
          \additemR\markup{#1}\to\@footnotelist
          \global\let\@footnotelist=\@footnotelist
     \fi
     \@footsf}

\def\footnotetext{%
     \ifx\@footnotelist\empty\err@extrafootnotetext\else\@footnotetext\fi}
\def\@footnotetext{%
     \getitemL\@footnotelist\to\@@foota
     \global\let\@footnotelist=\@footnotelist
     \insert\footins\bgroup
     \footnoteformat
     \splittopskip=\ht\strutbox\splitmaxdepth=\dp\strutbox
     \interlinepenalty=\interfootnotelinepenalty\floatingpenalty=\@MM
     \noindent\llap{\@@foota}\strut
     \bgroup\aftergroup\@footnoteend
     \let\@@scratcha=}
\def\@footnoteend{\strut\par\vskip\footnoteskip\egroup}

\def\footnoterule{\normalfonts
     \kern-.3em \hrule width2in height.04em \kern .26em }

\def\footnotecheck{%
     \ifx\@footnotelist\empty
     \else\err@extrafootnotemark
          \global\let\@footnotelist=\empty
     \fi}

\message{labels,}

\let\@@labeldef=\xdef
\newif\if@labelfile
\newwrite\@labelfile
\let\@prelabellist=\empty

\def\label#1#2{\trim#1\to\@@labarg\edef\@@labtext{#2}%
     \edef\@@labname{lab@\@@labarg}%
     \useafter\ifundefined\@@labname\then\else\@yeslab\fi
     \useafter\@@labeldef\@@labname{#2}%
     \ifstreaming
          \expandafter\toks@\expandafter\expandafter\expandafter
               {\csname\@@labname\endcsname}%
          \immediate\write\streamout{\noexpand\label{\@@labarg}{\the\toks@}}%
     \fi}
\def\@yeslab{%
     \useafter\ifundefined{if\@@labname}\then
          \err@labelredef\@@labarg
     \else\useif{if\@@labname}\then
               \err@labelredef\@@labarg
          \else\global\usename{\@@labname true}%
               \useafter\ifundefined{pre\@@labname}\then
               \else\useafter\ifx{pre\@@labname}\@@labtext
                    \else\err@badlabelmatch\@@labarg
                    \fi
               \fi
               \if@labelfile
               \else\global\@labelfiletrue
                    \immediate\write\sixt@@n{--> Creating file \jobname.lab}%
                    \immediate\openout\@labelfile=\jobname.lab
               \fi
               \immediate\write\@labelfile
                    {\noexpand\prelabel{\@@labarg}{\@@labtext}}%
          \fi
     \fi}

\def\putlab#1{\trim#1\to\@@labarg\edef\@@labname{lab@\@@labarg}%
     \useafter\ifundefined\@@labname\then\@nolab\else\usename\@@labname\fi}
\def\@nolab{%
     \useafter\ifundefined{pre\@@labname}\then
          \undefinedlabelformat
          \err@needlabel\@@labarg
          \useafter\xdef\@@labname{\undefinedlabelformat}%
     \else\usename{pre\@@labname}%
          \useafter\xdef\@@labname{\usename{pre\@@labname}}%
     \fi
     \useafter\newif{if\@@labname}%
     \expandafter\additemR\@@labarg\to\@prelabellist}

\def\prelabel#1{\useafter\gdef{prelab@#1}}

\def\ifundefinedlabel#1\then{%
     \expandafter\ifx\csname lab@#1\endcsname\relax}
\def\useiflab#1\then{\csname iflab@#1\endcsname}

\def\prelabelcheck{{%
     \def\^^\##1{\useiflab{##1}\then\else\err@undefinedlabel{##1}\fi}%
     \@prelabellist}}

\message{equation numbering,}

\newcount\chapternum
\newcount\sectionnum
\newcount\subsectionnum
\newcount\equationnum
\newcount\subequationnum
\newcount\figurenum
\newcount\subfigurenum
\newcount\tablenum
\newcount\subtablenum

\newif\if@subeqncount
\newif\if@subfigcount
\newif\if@subtblcount

\def\newchapternum{\newsectionnum=\z@\@resetnum\chapternum}
\def\newsectionnum{\newsubsectionnum=\z@\@resetnum\sectionnum}
\def\newsubsectionnum{\newequationnum=\z@\newfigurenum=\z@\newtablenum=\z@
     \@resetnum\subsectionnum}
\def\newequationnum{\newsubequationnum=\z@\@resetnum\equationnum}
\def\newsubequationnum{\@resetnum\subequationnum}
\def\newfigurenum{\newsubfigurenum=\z@\@resetnum\figurenum}
\def\newsubfigurenum{\@resetnum\subfigurenum}
\def\newtablenum{\newsubtablenum=\z@\@resetnum\tablenum}
\def\newsubtablenum{\@resetnum\subtablenum}

\def\@resetnum#1{\global\advance#1by1 \edef\next{\the#1\relax}\global#1}

\newchapternum=0

\def\chapternumstyle#1{\@setnumstyle\chapternum{#1}}
\def\sectionnumstyle#1{\@setnumstyle\sectionnum{#1}}
\def\subsectionnumstyle#1{\@setnumstyle\subsectionnum{#1}}
\def\equationnumstyle#1{\@setnumstyle\equationnum{#1}}
\def\subequationnumstyle#1{\@setnumstyle\subequationnum{#1}%
     \ifblank\subequationnumstyle\then\global\@subeqncountfalse\fi
     \ignorespaces}
\def\figurenumstyle#1{\@setnumstyle\figurenum{#1}}
\def\subfigurenumstyle#1{\@setnumstyle\subfigurenum{#1}%
     \ifblank\subfigurenumstyle\then\global\@subfigcountfalse\fi
     \ignorespaces}
\def\tablenumstyle#1{\@setnumstyle\tablenum{#1}}
\def\subtablenumstyle#1{\@setnumstyle\subtablenum{#1}%
     \ifblank\subtablenumstyle\then\global\@subtblcountfalse\fi
     \ignorespaces}

\def\eqnlabel#1{%
     \if@subeqncount
          \newsubequationnum=\next
     \else\newequationnum=\next
          \ifblank\subequationnumstyle\then
          \else\global\@subeqncounttrue
               \newsubequationnum=\@ne
          \fi
     \fi
     \label{#1}{\puteqnformat}(\puteqn{#1})%
     \ifdraft\rlap{\hskip.1in{\tt#1}}\fi}

\let\puteqn=\putlab

\def\equation#1#2{\useafter\gdef{eqn@#1}{#2\eqno\eqnlabel{#1}}}
\def\Equation#1{\useafter\gdef{eqn@#1}}

\def\putequation#1{\useafter\ifundefined{eqn@#1}\then
     \err@undefinedeqn{#1}\else\usename{eqn@#1}\fi}

\def\eqnseriesstyle#1{\gdef\@eqnseriesstyle{#1}}
\def\begineqnseries{\subequationnumstyle{\@eqnseriesstyle}%
     \defaultoption[]\@begineqnseries}
\def\@begineqnseries[#1]{\edef\@@eqnname{#1}}
\def\endeqnseries{\subequationnumstyle{blank}%
     \expandafter\ifnoarg\@@eqnname\then
     \else\label\@@eqnname{\puteqnformat}%
     \fi
     \aftergroup\ignorespaces}

\def\figlabel#1{%
     \if@subfigcount
          \newsubfigurenum=\next
     \else\newfigurenum=\next
          \ifblank\subfigurenumstyle\then
          \else\global\@subfigcounttrue
               \newsubfigurenum=\@ne
          \fi
     \fi
     \label{#1}{\putfigformat}\putfig{#1}%
     {\def\marginnoteformat{\tt}\marginnote{#1}}}

\let\putfig=\putlab

\def\figseriesstyle#1{\gdef\@figseriesstyle{#1}}
\def\beginfigseries{\subfigurenumstyle{\@figseriesstyle}%
     \defaultoption[]\@beginfigseries}
\def\@beginfigseries[#1]{\edef\@@figname{#1}}
\def\endfigseries{\subfigurenumstyle{blank}%
     \expandafter\ifnoarg\@@figname\then
     \else\label\@@figname{\putfigformat}%
     \fi
     \aftergroup\ignorespaces}

\def\tbllabel#1{%
     \if@subtblcount
          \newsubtablenum=\next
     \else\newtablenum=\next
          \ifblank\subtablenumstyle\then
          \else\global\@subtblcounttrue
               \newsubtablenum=\@ne
          \fi
     \fi
     \label{#1}{\puttblformat}\puttbl{#1}%
     {\def\marginnoteformat{\tt}\marginnote{#1}}}

\let\puttbl=\putlab

\def\tblseriesstyle#1{\gdef\@tblseriesstyle{#1}}
\def\begintblseries{\subtablenumstyle{\@tblseriesstyle}%
     \defaultoption[]\@begintblseries}
\def\@begintblseries[#1]{\edef\@@tblname{#1}}
\def\endtblseries{\subtablenumstyle{blank}%
     \expandafter\ifnoarg\@@tblname\then
     \else\label\@@tblname{\puttblformat}%
     \fi
     \aftergroup\ignorespaces}

\message{reference numbering,}

\newcount\referencenum \referencenum=0
\newcount\@@prerefcount \@@prerefcount=0
\newcount\@@thisref
\newcount\@@lastref
\newcount\@@loopref
\newcount\@@refseq
\newdimen\refnumindent
\let\@undefreflist=\empty

\def\referencenumstyle#1{\@setnumstyle\referencenum{#1}}

\def\referencestyle#1{\usename{@ref#1}}

\def\@refsequential{%
     \gdef\@refpredef##1{\global\advance\referencenum by\@ne
          \let\^^\=0\label{##1}{\^^\{\the\referencenum}}%
          \useafter\gdef{ref@\the\referencenum}{{##1}{\undefinedlabelformat}}}%
     \gdef\@reference##1##2{%
          \ifundefinedlabel##1\then
          \else\def\^^\####1{\global\@@thisref=####1\relax}\putlab{##1}%
               \useafter\gdef{ref@\the\@@thisref}{{##1}{##2}}%
          \fi}%
     \gdef\endputreferences{%
          \loop\ifnum\@@loopref<\referencenum
                    \advance\@@loopref by\@ne
                    \expandafter\expandafter\expandafter\@printreference
                         \csname ref@\the\@@loopref\endcsname
          \repeat
          \par}}

\def\@refpreordered{%
     \gdef\@refpredef##1{\global\advance\referencenum by\@ne
          \additemR##1\to\@undefreflist}%
     \gdef\@reference##1##2{%
          \ifundefinedlabel##1\then
          \else\global\advance\@@loopref by\@ne
               {\let\^^\=0\label{##1}{\^^\{\the\@@loopref}}}%
               \@printreference{##1}{##2}%
          \fi}
     \gdef\endputreferences{%
          \def\^^\####1{\useiflab{####1}\then
               \else\reference{####1}{\undefinedlabelformat}\fi}%
          \@undefreflist
          \par}}

\def\beginprereferences{\par
     \def\reference##1##2{\global\advance\referencenum by1\@ne
          \let\^^\=0\label{##1}{\^^\{\the\referencenum}}%
          \useafter\gdef{ref@\the\referencenum}{{##1}{##2}}}}
\def\endprereferences{\global\@@prerefcount=\the\referencenum\par}

\def\beginputreferences{\par
     \refnumindent=\z@\@@loopref=\z@
     \loop\ifnum\@@loopref<\referencenum
               \advance\@@loopref by\@ne
               \setbox\z@=\hbox{\referencenum=\@@loopref
                    \referencenumformat\enskip}%
               \ifdim\wd\z@>\refnumindent\refnumindent=\wd\z@\fi
     \repeat
     \putreferenceformat
     \@@loopref=\z@
     \loop\ifnum\@@loopref<\@@prerefcount
               \advance\@@loopref by\@ne
               \expandafter\expandafter\expandafter\@printreference
                    \csname ref@\the\@@loopref\endcsname
     \repeat
     \let\reference=\@reference}

\def\@printreference#1#2{\ifx#2\undefinedlabelformat\err@undefinedref{#1}\fi
     \noindent\ifdraft\rlap{\hskip\hsize\hskip.1in \tt#1}\fi
     \llap{\referencenum=\@@loopref\referencenumformat\enskip}#2\par}

\def\reference#1#2{{\par\refnumindent=\z@\putreferenceformat\noindent#2\par}}

\def\putref#1{\trim#1\to\@@refarg
     \expandafter\ifnoarg\@@refarg\then
          \toks@={\relax}%
     \else\@@lastref=-\@m\def\@@refsep{}\def\@more{\@nextref}%
          \toks@={\@nextref#1,,}%
     \fi\the\toks@}
\def\@nextref#1,{\trim#1\to\@@refarg
     \expandafter\ifnoarg\@@refarg\then
          \let\@more=\relax
     \else\ifundefinedlabel\@@refarg\then
               \expandafter\@refpredef\expandafter{\@@refarg}%
          \fi
          \def\^^\##1{\global\@@thisref=##1\relax}%
          \global\@@thisref=\m@ne
          \setbox\z@=\hbox{\putlab\@@refarg}%
     \fi
     \advance\@@lastref by\@ne
     \ifnum\@@lastref=\@@thisref\advance\@@refseq by\@ne\else\@@refseq=\@ne\fi
     \ifnum\@@lastref<\z@
     \else\ifnum\@@refseq<\thr@@
               \@@refsep\def\@@refsep{,}%
               \ifnum\@@lastref>\z@
                    \advance\@@lastref by\m@ne
                    {\referencenum=\@@lastref\putrefformat}%
               \else\undefinedlabelformat
               \fi
          \else\def\@@refsep{--}%
          \fi
     \fi
     \@@lastref=\@@thisref
     \@more}

\message{streaming,}

\newif\ifstreaming

\def\streamto{\defaultoption[\jobname]\@streamto}
\def\@streamto[#1]{\global\streamingtrue
     \immediate\write\sixt@@n{--> Streaming to #1.str}%
     \newwrite\streamout\immediate\openout\streamout=#1.str }

\def\streamfrom{\defaultoption[\jobname]\@streamfrom}
\def\@streamfrom[#1]{\newread\streamin\openin\streamin=#1.str
     \ifeof\streamin
          \expandafter\err@nostream\expandafter{#1.str}%
     \else\immediate\write\sixt@@n{--> Streaming from #1.str}%
          \let\@@labeldef=\gdef
          \ifstreaming
               \edef\@elc{\endlinechar=\the\endlinechar}%
               \endlinechar=\m@ne
               \loop\read\streamin to\@@scratcha
                    \ifeof\streamin
                         \streamingfalse
                    \else\toks@=\expandafter{\@@scratcha}%
                         \immediate\write\streamout{\the\toks@}%
                    \fi
                    \ifstreaming
               \repeat
               \@elc
               \input #1.str
               \streamingtrue
          \else\input #1.str
          \fi
          \let\@@labeldef=\xdef
     \fi}

\def\streamcheck{\ifstreaming
     \immediate\write\streamout{\pagenum=\the\pagenum}%
     \immediate\write\streamout{\footnotenum=\the\footnotenum}%
     \immediate\write\streamout{\referencenum=\the\referencenum}%
     \immediate\write\streamout{\chapternum=\the\chapternum}%
     \immediate\write\streamout{\sectionnum=\the\sectionnum}%
     \immediate\write\streamout{\subsectionnum=\the\subsectionnum}%
     \immediate\write\streamout{\equationnum=\the\equationnum}%
     \immediate\write\streamout{\subequationnum=\the\subequationnum}%
     \immediate\write\streamout{\figurenum=\the\figurenum}%
     \immediate\write\streamout{\subfigurenum=\the\subfigurenum}%
     \immediate\write\streamout{\tablenum=\the\tablenum}%
     \immediate\write\streamout{\subtablenum=\the\subtablenum}%
     \immediate\closeout\streamout
     \fi}


\def\err@badtypesize{%
     \errhelp={The limited availability of certain fonts requires^^J%
          that the base type size be 10pt, 12pt, or 14pt.^^J}%
     \errmessage{--> Illegal base type size}}

\def\err@badsizechange{\immediate\write\sixt@@n
     {--> Size change not allowed in math mode, ignored}}

\def\err@sizetoolarge#1{\immediate\write\sixt@@n
     {--> \noexpand#1 too big, substituting HUGE}}

\def\err@sizenotavailable#1{\immediate\write\sixt@@n
     {--> Size not available, \noexpand#1 ignored}}

\def\err@fontnotavailable#1{\immediate\write\sixt@@n
     {--> Font not available, \noexpand#1 ignored}}

\def\err@sltoit{\immediate\write\sixt@@n
     {--> Style \noexpand\sl not available, substituting \noexpand\it}%
     \it}

\def\err@bfstobf{\immediate\write\sixt@@n
     {--> Style \noexpand\bfs not available, substituting \noexpand\bf}%
     \bf}

\def\err@badgroup#1#2{%
     \errhelp={The block you have just tried to close was not the one^^J%
          most recently opened.^^J}%
     \errmessage{--> \noexpand\end{#1} doesn't match \noexpand\begin{#2}}}

\def\err@badcountervalue#1{\immediate\write\sixt@@n
     {--> Counter (#1) out of bounds}}

\def\err@extrafootnotemark{\immediate\write\sixt@@n
     {--> \noexpand\footnotemark command
          has no corresponding \noexpand\footnotetext}}

\def\err@extrafootnotetext{%
     \errhelp{You have given a \noexpand\footnotetext command without first
          specifying^^Ja \noexpand\footnotemark.^^J}%
     \errmessage{--> \noexpand\footnotetext command has no corresponding
          \noexpand\footnotemark}}

\def\err@labelredef#1{\immediate\write\sixt@@n
     {--> Label "#1" redefined}}

\def\err@badlabelmatch#1{\immediate\write\sixt@@n
     {--> Definition of label "#1" doesn't match value in \jobname.lab}}

\def\err@needlabel#1{\immediate\write\sixt@@n
     {--> Label "#1" cited before its definition}}

\def\err@undefinedlabel#1{\immediate\write\sixt@@n
     {--> Label "#1" cited but never defined}}

\def\err@undefinedeqn#1{\immediate\write\sixt@@n
     {--> Equation "#1" not defined}}

\def\err@undefinedref#1{\immediate\write\sixt@@n
     {--> Reference "#1" not defined}}

\def\err@nostream#1{%
     \errhelp={You have tried to input a stream file that doesn't exist.^^J}%
     \errmessage{--> Stream file #1 not found}}

\message{jyTeX initialization}

\everyjob{\immediate\write16{--> jyTeX version \fmtversion}%
     \edef\@@jobname{\jobname}%
     \edef\jobname{\@@jobname}%
     \settime
     \openin0=\jobname.lab
     \ifeof0
     \else\closein0
          \immediate\write16{--> Getting labels from file \jobname.lab}%
          \input\jobname.lab
     \fi}


\def\fixedskipslist{%
     \^^\{\topskip}%
     \^^\{\splittopskip}%
     \^^\{\maxdepth}%
     \^^\{\skip\topins}%
     \^^\{\skip\footins}%
     \^^\{\headskip}%
     \^^\{\footskip}}

\def\scalingskipslist{%
     \^^\{\p@renwd}%
     \^^\{\delimitershortfall}%
     \^^\{\nulldelimiterspace}%
     \^^\{\scriptspace}%
     \^^\{\jot}%
     \^^\{\normalbaselineskip}%
     \^^\{\normallineskip}%
     \^^\{\normallineskiplimit}%
     \^^\{\baselineskip}%
     \^^\{\lineskip}%
     \^^\{\lineskiplimit}%
     \^^\{\bigskipamount}%
     \^^\{\medskipamount}%
     \^^\{\smallskipamount}%
     \^^\{\parskip}%
     \^^\{\parindent}%
     \^^\{\abovedisplayskip}%
     \^^\{\belowdisplayskip}%
     \^^\{\abovedisplayshortskip}%
     \^^\{\belowdisplayshortskip}%
     \^^\{\abovechapterskip}%
     \^^\{\belowchapterskip}%
     \^^\{\abovesectionskip}%
     \^^\{\belowsectionskip}%
     \^^\{\abovesubsectionskip}%
     \^^\{\belowsubsectionskip}}


\def\twoupsetup{
     \topmargin=.75in
     \leftmargin=.5in
     \vsize=6.9in
     \hsize=4.75in
     \fullhsize=10in
     \let\draft=\relax}

\outputstyle{normal}                             

\def\marginnoteformat{\subscriptsize             
     \hsize=1in \baselinestretch=1000 \everypar={}%
     \tolerance=5000 \hbadness=5000 \parskip=0pt \parindent=0pt
     \leftskip=0pt \rightskip=0pt \raggedright}

\head={\ifdraft\normalfonts\it\hfil DRAFT\hfil   
     \llap{\number\day\ \monthword\month\ \militarytime}\else\hfil\fi}
\foot={\hfil\normalfonts\numstyle\pagenum\hfil}  

\normalbaselineskip=12pt                         
\normallineskip=0pt                              
\normallineskiplimit=0pt                         
\normalbaselines                                 

\topskip=.85\baselineskip
\splittopskip=\topskip
\headskip=2\baselineskip
\footskip=\headskip

\pagenumstyle{arabic}                            

\parskip=0pt                                     
\parindent=20pt                                  

\baselinestretch=1000                            


\chapterstyle{left}                              
\chapternumstyle{blank}                          
\def\chapterbreak{\newpage}                      
\abovechapterskip=0pt                            
\belowchapterskip=1.5\baselineskip               
     plus.38\baselineskip minus.38\baselineskip
\def\chapternumformat{\numstyle\chapternum.}     

\sectionstyle{left}                              
\sectionnumstyle{blank}                          
\def\sectionbreak{\vskip0pt plus4\baselineskip\penalty-100
     \vskip0pt plus-4\baselineskip}              
\abovesectionskip=1.5\baselineskip               
     plus.38\baselineskip minus.38\baselineskip
\belowsectionskip=\the\baselineskip              
     plus.25\baselineskip minus.25\baselineskip
\def\sectionnumformat{
     \ifblank\chapternumstyle\then\else\numstyle\chapternum.\fi
     \numstyle\sectionnum.}

\subsectionstyle{left}                           
\subsectionnumstyle{blank}                       
\def\subsectionbreak{\vskip0pt plus4\baselineskip\penalty-100
     \vskip0pt plus-4\baselineskip}              
\abovesubsectionskip=\the\baselineskip           
     plus.25\baselineskip minus.25\baselineskip
\belowsubsectionskip=.75\baselineskip            
     plus.19\baselineskip minus.19\baselineskip
\def\subsectionnumformat{
     \ifblank\chapternumstyle\then\else\numstyle\chapternum.\fi
     \ifblank\sectionnumstyle\then\else\numstyle\sectionnum.\fi
     \numstyle\subsectionnum.}


\footnotenumstyle{symbols}                       
\footnoteskip=0pt                                
\def\footnotenumformat{\numstyle\footnotenum}    
\def\footnoteformat{\footnotesize                
     \everypar={}\parskip=0pt \parfillskip=0pt plus1fil
     \leftskip=1em \rightskip=0pt
     \spaceskip=0pt \xspaceskip=0pt
     \def\\{\ifhmode\ifnum\lastpenalty=-10000
          \else\hfil\penalty-10000 \fi\fi\ignorespaces}}


\def\undefinedlabelformat{$\bullet$}             


\equationnumstyle{arabic}                        
\subequationnumstyle{blank}                      
\figurenumstyle{arabic}                          
\subfigurenumstyle{blank}                        
\tablenumstyle{arabic}                           
\subtablenumstyle{blank}                         

\eqnseriesstyle{alphabetic}                      
\figseriesstyle{alphabetic}                      
\tblseriesstyle{alphabetic}                      

\def\puteqnformat{\hbox{
     \ifblank\chapternumstyle\then\else\numstyle\chapternum.\fi
     \ifblank\sectionnumstyle\then\else\numstyle\sectionnum.\fi
     \ifblank\subsectionnumstyle\then\else\numstyle\subsectionnum.\fi
     \numstyle\equationnum
     \numstyle\subequationnum}}
\def\putfigformat{\hbox{
     \ifblank\chapternumstyle\then\else\numstyle\chapternum.\fi
     \ifblank\sectionnumstyle\then\else\numstyle\sectionnum.\fi
     \ifblank\subsectionnumstyle\then\else\numstyle\subsectionnum.\fi
     \numstyle\figurenum
     \numstyle\subfigurenum}}
\def\puttblformat{\hbox{
     \ifblank\chapternumstyle\then\else\numstyle\chapternum.\fi
     \ifblank\sectionnumstyle\then\else\numstyle\sectionnum.\fi
     \ifblank\subsectionnumstyle\then\else\numstyle\subsectionnum.\fi
     \numstyle\tablenum
     \numstyle\subtablenum}}


\referencestyle{sequential}                      
\referencenumstyle{arabic}                       
\def\putrefformat{\numstyle\referencenum}        
\def\referencenumformat{\numstyle\referencenum.} 
\def\putreferenceformat{
     \everypar={\hangindent=1em \hangafter=1 }%
     \def\\{\hfil\break\null\hskip-1em \ignorespaces}%
     \leftskip=\refnumindent\parindent=0pt \interlinepenalty=1000 }


\normalsize


\def\fmtversion{2.6M (June 1992)}

\catcode`\@=12
\def\upref#1/{\markup{[\putref{#1}]}}

\typesize=12pt

\footnoteskip=2pt
\footnotenumstyle{arabic}
%
%
\def\JSP#1#2#3{{\sl J. Stat. Phys.} {\bf #1} (#2) #3}
\def\PRL#1#2#3{{\sl Phys. Rev. Lett.} {\bf#1} (#2) #3}

\def\PTP#1#2#3{{\sl Prog. Theor. Phys.} {\bf #1} (#2) #3}
\def\ANYAS#1#2#3{{\sl Ann. New York Acad. Sci.} {\bf #1} (#2) #3}

\def\JPC#1#2#3{{\sl J. Physics} {\bf C#1} (#2) #3}
\def\JSM#1#2#3{{\sl J. Soviet Math.} {\bf #1} (#2) #3}
\def\JETP#1#2#3{{\sl Sov. Phys. JETP} {\bf #1} (#2) #3}
\def\sci#1#2#3{{\sl Science} {\bf #1} (#2) #3}
\def\PLA#1#2#3{{\sl Phys. Lett.} {\bf #1A} (#2) #3}
\def\JPCM#1#2#3{{\sl J. Physics Cond. Mat.} {\bf #1} (#2) #3}
\def\SSC#1#2#3{{\sl Solid State Comm.} {\bf #1} (#2) #3}
\def\TMP#1#2#3{{\sl Theor. Mat. Phys.} {\bf #1} (#2) #3}
\def\MPLB#1#2#3{{\sl Mod. Phys. Lett.} {\bf B#1} (#2) #3}
\def\NPB#1#2#3{{\sl Nucl. Phys.} {\bf B#1} (#2) #3}

\baselineskip=16pt
\def\pl#1{(\putlab{#1})}
\def\da{\downarrow}
\def\up{\uparrow}
\def\ga{\alpha}
\def\gd{\delta}
\def\ge{\epsilon}
\def\gs{\sigma}
\def\go{\omega}
\def\cd{c^\dagger}
\def\la{\lambda}
\def\p2{{\pi\over 2}}
\def\half{{1\over 2}}
\def\frac#1#2{{#1\over #2}}
\def\lh#1{\Lambda^h_{#1}}
\def\kh#1{{\rm sin}(k^h_{#1})}

\equation{H}{H = -\sum_{j=1}^L\sum_{\gs=\up ,\da}
\left(\cd_{j,\gs} c_{j+1,\gs} + \cd_{j+1,\gs} c_{j,\gs}\right)+
4U\sum_{j=1}^L (n_{j,\up}-{1\over 2})(n_{j,\da}-{1\over 2})\ .}

\equation{spin}{
S = \sum_{j=1}^L c^\dagger_{j,\up} c_{j,\da}\ , \quad
S^\dagger= \sum_{j=1}^L c^\dagger_{j,\da} c_{j,\up}\ ,\quad
S^z = \sum_{j=1}^L {1\over 2} (n_{j,\da}-n_{j,\up}) \ ,}

\equation{eta}{
\eta = \sum_{j=1}^L (-1)^jc_{j,\up} c_{j,\da}\ , \quad
\eta^\dagger = \sum_{j=1}^L (-1)^jc^\dagger_{j,\da} c^\dagger_{j,\up}\ ,
\quad \eta^z = {1\over 2}\sum_{j=1}^L (n_{j,\up}+n_{j,\da}  - 1)\ .}

\equation{eps}{\eqalign{
p_s(\la) &= \p2 - \int_{0}^\infty {d\go\over \go} {J_0(\go) {\rm
sin}(\go\la)\over {\rm cosh}(\go U)}\qquad 0<p_s(\la)<\pi\cr
\ge_s(\la) &= 2 \int_{0}^\infty {d\go\over \go} {J_1(\go) {\rm
cos}(\go\la)\over {\rm cosh}(\go U)}\ .\cr}}

\equation{epc}{\eqalign{
p_h(k^h) &= \p2-k^h - \int_{0}^\infty {d\go\over \go} {J_0(\go) {\rm
sin}(\go\ {\rm sin}(k^h))e^{-\go U}\over {\rm cosh}(\go U)}\cr
\ge_h(k^h) &= 2U+2\ {\rm cos}(k^h) + 2 \int_{0}^\infty {d\go\over \go}
{J_1(\go) {\rm cos}(\go\ {\rm sin}(k^h))e^{-\go U}\over {\rm cosh}(\go
U)}\ .\cr}}

\equation{antiholon}{\eqalign{
\ge_{ah}(k) &= \ge_h(k)\ ,\quad p_{ah}(k) = -\pi + p_h(k)\ .\cr}}

\equation{lambdastr}{
\La_{\alpha}^{m,j} = \La_{\alpha}^m - \frac{1}{2} (m+1-2j) u
\qquad \La_{\alpha}^m {\ \rm real}\qquad j=1,2, \ldots, m \ .}

\equation{klstr1}{\eqalign{
k_\alpha^1 &= \pi - \sin^{-1}(\La_\alpha^{\prime \  m} - \frac{1}{2} m u)
\cr
k_\alpha^2 &= \sin^{-1}(\La_\alpha^{\prime \  m} - \frac{1}{2} (m-2) u)
\cr
k_\alpha^3 &= \pi - k_\alpha^2
\cr
k_\alpha^4 &= \sin^{-1}(\La_\alpha^{\prime \  m} - \frac{1}{2} (m-4) u)
\cr
k_\alpha^5 &= \pi - k_\alpha^4
\cr
\ldots
\cr
k_\alpha^{2m-2} &= \sin^{-1}(\La_\alpha^{\prime \  m} + \frac{1}{2} (m-2) u)
\cr
k_\alpha^{2m-1} &= \pi - k_\alpha^{2m-2}
\cr
k_\alpha^{2m} &=
  \pi - \sin^{-1}(\La_\alpha^{\prime \  m} + \frac{1}{2} m u) \ .\cr}}

\equation{klstr2}{\La_{\alpha}^{\prime\ m,j} = \La_{\alpha}^{\prime\ m}
  - \frac{1}{2} (m+1-2j) u \ , \quad \La_{\alpha}^{\prime\ m}
{\ \rm real}\qquad j=1,2, \ldots, m \ . }

\equation{cons}{\
N_e = M_e + 2\  \sum_{m=1}^{\infty} m \  M^\prime_m \ ,
\hskip 6mm
M = \sum_{m=1}^{\infty} m (M_m + M_m^\prime) \ .}

\equation{pbc}{\eqalign{
& k_j L =
   2 \pi I_j
 - \sum_{n=1}^\infty \sum_{\alpha =1}^{M_n}
   \theta \left( \frac{\sin k_j - \Lambda_\alpha^n}{n U} \right)
 - \sum_{n=1}^\infty \sum_{\alpha =1}^{M^\prime_n}
   \theta \left( \frac{\sin k_j - \Lambda_\alpha^{\prime n}}
                      {n U} \right) \ ,\cr
& \sum_{j=1}^{N_e-2M^\prime}
   \theta \left( \frac{\Lambda_\alpha^n - \sin k_j}
                      {n U} \right)
 = 2 \pi  J_\alpha^n
   + \sum_{m=1}^\infty\sum_{\beta =1}^{M_m} \Theta_{nm} \left(
     \frac{\Lambda_\alpha^n - \Lambda_\beta^m} {U} \right) \ ,\cr}}
\equation{pbc2}{\eqalign{
& L \left[ sin^{-1} \left( \Lambda_\alpha^{\prime n} + i n U \right)
  + sin^{-1} \left( \Lambda_\alpha^{\prime n} - i n U
\right) \right]\cr
& \hskip 1cm =
   2 \pi J_\alpha^{\prime n}
 + \sum_{j=1}^{N_e-2M^\prime}
   \theta \left( \frac{\Lambda_\alpha^{\prime n} - \sin k_j}
                      {n U} \right)
 + \sum_{m=1}^\infty\sum_{\beta =1}^{M_m^\prime} \Theta_{nm} \left(
     \frac{\Lambda_\alpha^{\prime n} - \Lambda_\beta^{\prime m}}
                      {U} \right) \ ,\cr}}

\equation{theta}{\theta_{nm}(x) = \cases{\theta ({x\over{|n-m|}}) +
2\ \theta ({x\over{|n-m|+2}})+\dots +2\ \theta ({x\over{n+m-2}}) +
\theta ({x\over{n+m}})&if $n\ne m$\cr\cr
2\ \theta ({x\over{2}}) + 2\ \theta ({x\over{4}})+\dots +
2\ \theta ({x\over{2n-2}})+\theta ({x\over{2n}})&if $n=m$\ \ .\cr}}

\equation{ineq}{\eqalign{
\mid J_\alpha^n \mid \leq & \half (N_e - 2 M^\prime
     - \sum_{m=1}^\infty t_{nm} M_m -1) \ ,\cr
\mid J_\alpha^{\prime n} \mid \leq & \half (L - N_e + 2 M^\prime
     - \sum_{m=1}^\infty t_{nm} M_m^\prime -1) \ ,\ -{L-1\over 2}\leq
I_j \leq {L-1\over 2}\ ,\cr}}

\equation{ss}{S(\lh{1},\lh{2}) = -{\Gamma({1+i\mu\over 2})
\Gamma(1-{i\mu\over 2}) \over \Gamma({1-i\mu\over 2})
\Gamma(1+{i\mu\over 2})}\ \left({\mu\over\mu+i} I + {i\over
\mu+i}\Pi\right)\ ,\ }

\equation{matrices}{I=\left(\matrix{1&0&0&0\cr 0&1&0&0\cr 0&0&1&0\cr
0&0&0&1\cr}\right)\ , \Pi=\left(\matrix{1&0&0&0\cr 0&0&1&0\cr 0&1&0&0\cr
0&0&0&1\cr}\right)\ .}

\equation{cc}{S(\kh{1}, \kh{2}) = {\Gamma({1-i\mu\over 2})
\Gamma(1+{i\mu\over 2}) \over \Gamma({1+i\mu\over 2})
\Gamma(1-{i\mu\over 2})}\ \left({\mu\over\mu-i} I - {i\over
\mu-i}\Pi\right)\ ,}

\equation{sc}{S(\lh{},\kh{}) = -i\ {1+i\ \exp({\pi}\mu)\over 1-i\
\exp({\pi}\mu)} I\ ,\ \mu={\lh{}-\kh{}\over 2U}>0\ .}

\equation{parthol}{c_{j,\up}\leftrightarrow (-1)^j c^\dagger_{j,\up}\
,}

\equation{aeps}{\eqalign{
p_c^p(\la) &= \pi - \int_{0}^\infty {d\go\over \go} {J_0(\go) {\rm
sin}(\go\la)\over {\rm cosh}(\go U)} = \pi - p_c^h(\la)\ ,\cr
\ge_c^p(\la) &= 2 \int_{0}^\infty {d\go\over \go} {J_1(\go) {\rm
cos}(\go\la)\over {\rm cosh}(\go U)}=\ge_c^h(\la) \ ,\cr}}

\equation{aepc}{\eqalign{
p_s(k) &= \pi -k - \int_{0}^\infty {d\go\over \go} {J_0(\go) {\rm
sin}(\go\ {\rm sin}(k))e^{-\go |U|}\over {\rm cosh}(\go U)}\cr
\ge_s(k) &= 2|U|+2\ {\rm cos}(k) + 2 \int_{0}^\infty {d\go\over \go}
{J_1(\go) {\rm cos}(\go\ {\rm sin}(k))e^{-\go |U|}\over {\rm cosh}(\go
U)}\ ,\cr}}

\pagenumstyle{blank}
\footnoteskip=2pt
\line{\it July 1993\hfil ITP-SB-93-40}
\vskip2em
\baselineskip=32pt
\begin{center}
{\bigsize{\sc Scattering Matrix and Excitation Spectrum\\
of the Hubbard Model}}
\end{center}
\vfil
\baselineskip=16pt

\begin{center}
{\bigsize
Fabian H.L.E\sharps ler\footnote[$\ \flat$]{\sc e-mail:
fabman@max.physics.sunysb.edu\\ \sc Address
after Sept.1, 1993: {\smallfonts\rm Physikalisches Institut der
Universit\"at Bonn, Nussallee 12, 53115 Bonn, Germany}}
\vskip .5cm
and \vskip .5cm
Vladimir E. Korepin\footnote[$\ \sharp$]{\sc e-mail:
korepin@max.physics.sunysb.edu}\vskip .5cm}

\vskip 0.7cm

\it Institute for Theoretical Physics\vskip 4pt
State University of New York at Stony Brook\vskip 4pt
Stony Brook, NY~~11794-3840
\end{center}

\vfil

\centertext{\bfs \bigsize ABSTRACT}
\vskip\belowsectionskip

\begin{narrow}[4em]
\noindent
We consider the one-dimensional Hubbard model at half filling.
We show that both excitation spectrum and S-matrix are determined by
the SO(4) symmetry of the model. The complete set of excitations
is given by the scattering states four elementary excitations, which
form the fundamental representation of SO(4). We evaluate the exact
S-matrix, which satisfies the Yang-Baxter relation. The results for
the repulsive and attractive Hubbard model are related by an
interchange of spin and charge degrees of freedom.
\end{narrow}
\vskip .5cm
{\sc Pacs: 75.10.Jm , 05.50.q}
\vfil

\break

\pagenumstyle{arabic}

Strongly correlated electronic systems are currently under intense
study in relation with high-$T_c$ superconductivity.
The two-dimensional Hubbard model is in the very center of these
investigations. It is believed to share important features with its
one-dimensional analog\upref pwa/.
A very interesting problem is the separation of spin and charge in the
Hubbard model in one\upref woynar1, woynar2/ and two dimensions\upref
pwa2/. In one spatial dimension the Hubbard model can be solved exactly
by means of the Bethe Ansatz\upref lieb/. The Hubbard hamiltonian is
given by the following expression

$$\putequation{H}$$

Here $\cd_{j,\gs}$ are canonical fermionic creation operators on the
lattice, $j$ labels the sites of a chain of even length $L$, $\gs$
labels the spin degrees of freedom, $U$ is the coupling constant, and
$n_{j,\gs} = \cd_{j,\gs}c_{j,\gs}$ is the number operator for spin
$\gs$ on site $j$. The hamiltonian \pl{H} is invariant under a
SO(4)$=$SU(2)$\times$SU(2)/$Z_2$ algebra\upref hl,yang1,yang2/. The
two SU(2) algebras are the spin SU(2) with generators
$$\putequation{spin}$$
and the $\eta$-SU(2) algebra with generators
$$\putequation{eta}$$

All $6$ generators commute with the hamiltonian \pl{H}. Below we first
give a complete classification of the excitation spectrum of the {\sl
repulsive} Hubbard model by means of its SO(4) symmetry. The results
are such that the spin and charge degrees of freedom completely
separate from one another on the level of the elementary excitations.
We then determine the exact two-particle S-matrix, which is of
importance in the study of transport properties\upref Carmelo/. The
S-matrix is SO(4) invariant and satisfies the Yang-Baxter relation,
which implies the 2-particle reducibility of the N-body scattering
matrix. We conclude with a discussion of excitation spectrum and
S-matrix of the {\sl attractive} Hubbard model. Detailed
derivations of the results presented in this letter can be found in
[\putref{sm}].\vskip .4cm


Let us now turn to a classification of the excitation spectrum over the
repulsive half-filled ground state ($U>0$). In the many previous works
on this problem (most notably [\putref{woynar1,woynar2,ov}]) the
correct dispersions for many excitations were obtained. However, the
Lie-algebraic structure of the excitation spectrum was not fully
revealed. We classify {\sl all} excitations with respect to
representations of the SO(4) symmetry algebra.
The spectrum is described by the scattering states of four
elementary excitations only. Two of the elementary excitations carry
spin but no charge. This means that they transform in the spin-$1\over
2$ representation of the spin-SU(2), and are a singlet with respect to
the $\eta$-SU(2). The resulting SO(4)-representation is denoted by
$({1\over 2},0)$. We shall call these two excitations {\sl spinons}
(They are similar to the two elementary excitations in the
spin-$1\over 2$ Heisenberg XXX antiferromagnet at zero magnetic
field\upref ft,ft2,takh/). Energy and momentum of these particles are
expressed in terms of a spectral parameter $\la$. It plays the same
role as the rapidity for the case of relativistic particles. Energy
$\ge_s(\la)$ and  momentum $p_s(\la)$
$$\putequation{eps}$$
\noindent
Here $J_n(\go)$ are Bessel functions. This dispersion relation was
obtained in [\putref{woynar1, ov}].\vfil
The other two elementary excitations carry charge $\pm e$ but no spin.
They are called {\sl holon} and {\sl antiholon} and form the
$(0,{1\over 2})$-representation of SO(4), {\sl i.e.} they transform in
the spin-$1\over 2$ representation of the $\eta$-SU(2) \pl{eta} and
are a singlet with respect to the spin-SU(2)\pl{spin}.
The momentum of holon and antiholon differs by $\pi$ as $\eta^\dagger$
does not commute with the momentum operator but changes the momentum
by $\pi$. The holon carries charge $+e$, where $-e$ is the charge of
the electron. Energy and momentum are parametrized by the ``bare''
momentum $k^h$ of the holon, and ${\rm sin}(k^h)$ plays the role of
the spectral parameter. Energy and momentum are given by\upref
woynar2/
$$\putequation{epc}$$
The antiholon can be obtained from the wave function of the holon by
applying the operator $\eta^\dagger$. Its energy and momentum are
found to be
$$\putequation{antiholon}$$

The spectrum of all excitations of the repulsive half-filled Hubbard
model is given by the {\sl scattering states} of the four elementary
excitations listed above.
The proof of this is similar to the proof of the corresponding
statement in the spin-$1\over 2$ Heisenberg XXX antiferromagnet\upref
ft,ft2,takh/. The basis of the proof is a theorem of [\putref{eks}],
which states that (i) All Bethe Ansatz states with $N_\da<N_\up$ below
(or at) half-filling are lowest weight states of the SO(4) symmetry
algebra, and (ii) The set of states obtained by acting with the SO(4)
raising operators on the Bethe states is {\sl complete}, {\sl i.e.} it
forms a basis of the electronic Hilbert space. This theorem implies
that a complete set of excitations over the ground state is obtained by
constructing excited states by means of the Bethe Ansatz and then act
on these with the SO(4) raising operators. A complete set of Bethe
Ansatz excitations can be constructed by using the zero-temperature
limit of the Thermodynamic Bethe Ansatz\upref taka/. Like in the
XXX-antiferromagnet\upref ft,ft2/ elementary excitations always occur
in {\sl pairs}. This is due to the $Z_2$-factor, which projects out
all representations of SU(2)$\times$SU(2) with half-odd integer
quantum number: $S^z+\eta^z = N_\da + {L\over 2}$ ($L$ is even).
Accordingly there are no one-particle excitations, and more generally
no excitations with and {\sl odd} number of particles (spinons or
holons/antiholons) in the half-filled Hubbard model.
\vskip .4cm

In order to construct the two-particle scattering matrix a complete
knowledge of all excitations in the two-particle sector is necessary.
In the two-particle sector there exist three kinds of excitations:
Scattering states of two spinons (forming the representation
$(1,0)\oplus (0,0)$ ); Scattering states of two holons (in the
representation $(0,1)\oplus (0,0)$); and finally scattering states of
one holon and one spinon (forming the $({1\over 2},{1\over 2})$
representation of SO(4) ). Bethe states in the Hubbard model are
parametrized by sets of spectral parameters $\{ k_j\},\
\{\Lambda^n_\ga\},\ \{\Lambda^{\prime n}_\ga\}$, which are subject to
the following sets of coupled algebraic equations\upref taka/

$$\putequation{pbc}$$
$$\putequation{pbc2}$$

Here $L=2\times {\rm odd}$ is the even length of the lattice, $I_j$,
$J^n_\ga$, and $J^{\prime n}_\ga$ are integer or half-odd integer
numbers, $N_e$ is the total number of electrons,
$M^\prime=\sum_{n=1}^\infty M^\prime_n$, $\theta(x)=2{\rm arctan}(x)$,
and
$$\putequation{theta}$$
The spectral parameters $\Lambda^n_\ga$, $\Lambda^{\prime n}_\ga$ are
the real centers of complex roots (``strings'' of lenght $n$) of the
Lieb-Wu equations (see [\putref{taka}]), $M_n$ and $M^\prime_n$
are the numbers of $\Lambda$-strings of length $n$ and
$\Lambda^\prime$-strings of length $n$ in a specific solution of the
system \pl{pbc}. The integer (half-odd integer) numbers in \pl{pbc}
have the ranges \vfill

$$\putequation{ineq}$$
where $t_{mn} = 2{\rm min}\{m,n\}-\gd_{mn}$. Each set of ``integers''
$\{I_j\},\ \{J^n_\ga\},\ \{J^{\prime n}_\ga\}$ is in one-to-one
correspondence with a set of spectral parameters, which in turn
unambiguously specifies one eigenstate of the hamiltonian \pl{H}. Thus
the ground state and all excited states
can be constructed by specifying sets of ``integers'' in \pl{pbc} and
then taking the thermodynamic limit, in which \pl{pbc} reduce to
coupled integral equations (see {\sl e.g.} [\putref{lieb,woynar1,
woynar2,ov}]). The half-filled ground state is obtained by taking
$M_1={L\over 2}$ and $N_e=L$ in \pl{pbc} ($M_n=0\ \forall n\geq 2,\
M^\prime_n=0\ \forall n$) and filling all vacancies for the
``integers'' $I_j$ and $J^1_\ga$ symmetrically around zero\upref
lieb/. This amounts to filling the Fermi seas of spectral parameters
$k_j$ between $-\pi$ and $\pi$, and of $\Lambda^1_\ga$ between
$-\infty$ and $\infty$. The ground state is an SO(4) singlet
$(0,0)$.\vfil
The scattering states of two spinons are similar to the
the XXX-antiferromagnet\upref ft,ft2,takh/. The lowest weight state
of the spin-triplet ($S=1$, $S^z=-1$, $\eta = 0$) is obtained by
diminishing the number of down spins by one as compared to the ground
state $M_1 = {L\over 2}-1$. This enlarges the allowed range of
``integers'' $J^1_\ga$ \pl{ineq} by one, so that we end up with {\sl
two} holes in the Fermi sea of the $\Lambda^1_\ga$. We denote the
corresponding spectral parameters by $\lh{1}$ and $\lh{2}$. Energy and
momentum of this state are given in terms of the two spinons according
to $E_{ST} = \ge_s(\lh{1})+\ge_s(\lh{2})$ and $P_{ST} =
p_s(\lh{1})+p_s(\lh{2})$.
The lowest weight theorem cited above guarantees that not only
$S^z=-1$ but that also $S=1$. The two other states of the triplet can
be obtained by application of $S^\dagger$.\\
The $S=0, \eta =0$ spin-singlet scattering state of two spinons is
constructed by choosing $M_1 = {L\over 2}-2$ and $M_2=1$. This
introduces again two holes in the $\Lambda^1_\ga$-distribution. In
addition there is one $\Lambda$-string of lenght $2$ present. The
allowed range of ``integers'' for the $2$-string \pl{ineq} is $0$
only. Solving the equation \pl{pbc} for the spectral parameter
$\Lambda^{(2)}$ we find $\Lambda^{(2)}={\lh{1} +\lh{2}\over 2}$.
Energy and momentum of the singlet are the same as for the triplet.\\
The scattering state of a spinon (with spin up) and a holon can be
constructed from the Bethe Ansatz by choosing $N_e=L-1$ and
$M_1={\L\over 2}-1$, which introduces one hole in the distribution of
the $k_j$'s and the $\Lambda^1_\ga$'s respectively. We denote the
corresponding spectral parameters by $k^h$ and $\lh{}$. The spin
of this excitation is $S={1\over 2}=\eta ,\ S^z=-{1\over 2}= \eta^z$.
Accordingly it carries charge $+e$. Energy and momentum are gven by
$E_{sc} = \ge_s(\lh{})+\ge_h(k^h)$ and $P_{sc} =
p_s(\lh{})+p_h(k^h)$. The other three states of the SO(4) multiplet
can be obtained by applying $S^\dagger$, $\eta^\dagger$, and both.\\
The scattering state of two holons
with $\eta = 1,\eta^z=-1, S=0$ (charge triplet)
is constructed by making two holes in the distribution of $k_j$'s
($N_e=L-2$). We denote the corresponding spectral parameters by
$k^h_1$ and $k^h_2$. Energy and momentum of this state are given by
$E_{CT} = \ge_h(k^h_1)+\ge_h(k^h_2)$ and $P_{CT} = p_h(k^h_1) +
p_h(k^h_2)$. The other two states of the multiplet are obtained by
applying $\eta^\dagger$.\\
The charge singlet $S=0,\eta =0=\eta^z$ scattering state is
characterized by taking $N_e=L$, $M_1={L\over 2}-1$
and $M^{\prime 1}=1$ (insertion of a bound state of two $k$'s and one
$\Lambda^\prime$). This introduces again two holes in the $k_j$ Fermi
sea, and in addition there is a so-called $\Lambda^\prime$-string
present. The spectral parameter of the $\Lambda^\prime$-string is
determined by the system \pl{pbc2} to be $\Lambda^{\prime 1} = {{\rm
sin}(k^h_1) + {\rm sin}(k^h_2) \over 2}$. Energy and momentum of the
$\eta$-singlet are $E_{CS} = \ge_{ah}(k^h_1)+\ge_h(k^h_2)$ and
$P_{CS} = p_{ah}(k^h_1) + p_h(k^h_2)$. This completes our
discription of all two-particle states.
\vskip .4cm

We now turn to the evaluation of the two-body scattering matrix. In
Quantum Mechanics the scattering matrix can be extracted from the
asymptotics of the wave-function of the scattering state\upref
Landau/. The scattering of the collective excitations described above
is a quantum mechanical problem. However, the ground state is highly
nontrivial, as are the many-body wave functions given by the Bethe
Ansatz. In [\putref{korepin}] it was explained how to modify the rules
of quantum mechanical scattering theory in order to obtain the correct
scattering matrix for Bethe Ansatz solvable models ([\putref{korepin}]
deals with the derivation of Zamolodchikov's S-matrix starting from
the Bethe Ansatz).  The same method has been applied to the case of
the XXX Heisenberg model\upref ft,ft2,takh/ (see also
[\putref{vladb}]). More precisely, the method of [\putref{korepin}]
permits the evaluation of the logarithmic derivative of the S-matrix,
and thus the S-matrix up to an overall constant factor. We have fixed
this factor by considering the infinite $U$ limit of holon-holon
scattering. In this limit the S-matrix must reduce to the one of free
fermions, which is equal to $1$. This allows us to solve the problem
of fixing the constant factor in the S-matrix of the
XXX-antiferromagnet: the S-matrix for that model is given by the
expression (11). The complete S-matrix for the Hubbard model is
$16\times 16$ dimensional and blockdiagonal. The block describing
scattering of two spinons (with spectral parameters $\lh{1}$ and
$\lh{2}$) is $4\times 4$ matrix of the form

$$\putequation{ss}$$
\noindent where $\mu={\lh{1}-\lh{2}\over 2U}$ and
$$\putequation{matrices}$$
The physical region is defined as $\mu>0$ (only then scattering takes
place). As a function of the variable $\mu$ the S-matrix \pl{ss}
coincides with the S-matrix of the spin-$1\over 2$ Heisenberg
antiferromagnet. This does not hold as a function of the momentum.
The $4\times 4$ block describing the scattering of two charged particles
(holons/antiholons) with spectral parameters ${\rm sin}(k^h_1)$ and
${\rm sin}(k^h_2)$ is

$$\putequation{cc}$$
\noindent where $\mu={\kh{1}-\kh{2}\over 2U}>0$.
The block describing the scattering of one spinon on one
holon/antiholon is purely diagonal

$$\putequation{sc}$$
The last block describes the scattering of one
holon/antiholon on one spinon and is given by the same expression
\pl{sc}, where now $\mu={\kh{}-\lh{}\over 2U}>0$.
The complete S-matrix satisfies the Yang-Baxter equation,
which implies that the many-body scattering matrix factorizes into the
product of two-body S-matrices. Clearly the S-matrix reflects the
SO(4) symmetry of the model. The analytical structure of the
S-matrix is as follows: The allowed strip for rapidities is
$|{\rm Im}(\lh{})|\le U$, and $|{\rm Im}(k^h)|\le 2U$. The S-matrix
has poles in the complex rapidity plane, but all of them are
unphysical and do {\sl not} correspond to bound states.

\vskip .4cm

The structure of elementary excitations in the attractive Hubbard
model is as follows: there is a total of four elementary excitations,
grouped into doublets of the spin and $\eta$-SU(2) respectively. The
chargeless spin-carriers have the dispersion
$$\putequation{aepc}$$
\noindent which up to the shift of momentum by $\pi\over 2$ is the
same as for the holon in the repulsive case.
The spinless charge-carriers (one particle, one hole) have the
dispersions\upref woynar/
$$\putequation{aeps}$$
which are (up to a constant in the momenta) the same as the spinon
dispersions in the repulsive case. The S-matrix in the attractive case
coincides with the one for the repulsive case {\sl as a function of
the unformising parameter $\mu$}. There also exists a naive argument
for the relation of attractive and repulsive Hubbard model by an
interchange of charge and spin degrees of freedom. It is a well known
fact that attractive and repulsive Hubbard model are connected by the
particle-hole transformation (for spin-up only)

$$\putequation{parthol}$$
\noindent under which the hamiltonian \pl{H} transforms according to
$H(U)\rightarrow H(-U)$. The transformation \pl{parthol} effectively
interchanges spin and charge degrees of freedom, {\sl e.g.} transforms
the spin-SU(2) \pl{spin} into the $\eta$-SU(2) \pl{eta} and
vice versa.

In this letter we classified the excitation spectrum and the
Scattering matrix of the one-dimensional Hubbard model by means of its
$SO(4)$ symmetry. This emphasizes the importance of symmetries in
systems of strongly correlated electrons. The S-matrix is an important
tool in the study of transport properties\upref Carmelo/. The
classification of elementary excitations into spinons (which carry
spin but no charge) and holons/antiholons (which carry charge but no
spin) gives a precise meaning to spin-and charge separation in one and
two-dimensional models of correlated electrons. In integrable models
the S-matrix is often proportional to the R-matrix. In
[\putref{shastry}] a R-matrix for an embedding of the Hubbard
hamiltonian in a commuting family of transfer matrices was derived.
This R-matrix, which belongs to a {\sl fundamental model} (where
L-operator and R-matrix are essentially identical\upref vladb/), is
different from our S-matrix. It would be interesting to search for a
different embedding of the Hubbard hamiltonian, using a SO(4)
invariant R-matrix proportional to our S-matrix. The L-operator for
such a construction would have to be different from the R-matrix.

\vskip .3cm
\centerline{\sc Acknowledgements:}
\vskip .2cm
It is a pleasure to thank Prof. C.N. Yang and Dr. E. Melzer for
stimulating discussions. This work was supported in part by NATO
9.15.02 RG 901098 Special Panel on Chaos Order and Patterns
`Functional Integral Methods in Statistical Mechanics and
Correlations'. \vfil
\begin{putreferences}
\centerline{{\sc References}}
\smallfonts
\vskip .5cm
\reference{pwa}{P.W. Anderson,\ \sci{235}{1987}{1196}.}
\reference{pwa2}{G. Baskaran, Z. Zou, P.W. Anderson,\
\SSC{63}{1987}{973}.}
\reference{lieb}{E.H. Lieb, F.Y. Wu,\ \PRL{20}{1968}{1445}.}
\reference{hl}{O.J. Heilmann, E.H. Lieb,\ \ANYAS{172}{1971}{583}.}
\reference{yang}{C.N. Yang,\ \PRL{19}{1967}{1312}.}
\reference{yang1}{C.N. Yang,\ \PRL{63}{1989}{2144}.}
\reference{yang2}{C.N. Yang and S. Zhang,\ \MPLB{4}{1990}{759}.}
\reference{ft}{L.D. Faddeev, L. Takhtajan, \JSM{24}{1984}{241}.}
\reference{ft2}{L.D. Faddeev, L. Takhtajan,\ \PLA{85}{1981}{375}.}
\reference{takh}{L. Takhtajan,\ \PLA{87}{1982}{479}.}
\reference{woynar}{F. Woynarovich,\ \JPC{16}{1983}{6593}.}
\reference{woynar1}{F. Woynarovich,\ \JPC{16}{1983}{5293}.}
\reference{woynar2}{F. Woynarovich,\ \JPC{15}{1982}{85}.}
\reference{vladb}{V.E. Korepin, G. Izergin and N.M. Bogoliubov,\ {\sl
Quantum Inverse Scattering Method, Correlation Functions and Algebraic
Bethe Ansatz}, Cambridge University Press, 1993}
\reference{korepin}{V.E. Korepin,\ \TMP{76}{1980}{165}.}
\reference{eks}{F.H.L. E\char'31ler, V.E. Korepin, K. Schoutens,\
\NPB{372}{1992}{559}, \NPB{384}{1992}{431}, \PRL{67}{1991}{3848}.}
\reference{Landau}{L.D. Landau, E.M. Lifshitz, ``{\sl Quantum
Mechanics}'', Pergamon Press 1975.}
\reference{ov}{A.A. Ovchinnikov,\ \JETP{30}{1970}{1160}.}
\reference{taka}{M. Takahashi,\ \PTP{47}{1972}{69}.}
\reference{sm}{F.H.L. E\char'31ler, V.E. Korepin,\ {\sl SO(4)
Invariant Scattering Matrix of the Hubbard Model, preprint}}
\reference{Carmelo}{J. Carmelo, A.A. Ovchinnikov,\ \JPCM{3}{1991}{757}.}
\reference{shastry}{B.S. Shastry,\ \JSP{50}{1988}{57}.}
\reference{rv}{{\sl Exactly Solvable Models of Strongly Correlated
Electrons}, eds V.E. Korepin, F.H.L. E\char'31ler,\ World Scientific,
{\sl to appear}}

\end{putreferences}
\bye